\documentclass[trackchanges,twocolumn]{aastex7}

\submitjournal{ApJ}

\usepackage{amsmath}
\usepackage{url}
\usepackage{comment}
\usepackage{ulem}

\received{August 5, 2025}
\revised{February 16, 2026}
\accepted{\today}

\graphicspath{{./}{figs/}}

\shortauthors{Ikuta et al.}

\defcitealias{Namekata23}{Paper~I}
\defcitealias{Namekata24}{Paper~I\hspace{-.1em}I}

\newcommand{\nasa}{NASA Goddard Space Flight Center, 8800 Greenbelt Road, Greenbelt, MD 20771, USA}

\begin{document}

\shorttitle{Starspot mapping and Multiwavelength Variability for a young solar-type star}

\title{Multiwavelength Campaign Observations of a Young Solar-type Star, EK Draconis. I\hspace{-.1em}I\hspace{-.1em}I. \\ Comparison between Starspot Mapping, Zeeman Doppler Imaging, and Multiwavelength Variability}

\correspondingauthor{Kai Ikuta}

\author[0000-0002-5978-057X]{Kai Ikuta}
\altaffiliation{Current Affiliation: Department of Social Data Science, Hitotsubashi University, 2-1 Naka, Kunitachi, Tokyo 186-8601, Japan}
\affil{Department of Multidisciplinary Sciences, The University of Tokyo, 3-8-1 Komaba, Meguro, Tokyo 153-8902, Japan}
\email[show]{kaiikuta.astron@gmail.com}

\author[0000-0002-1297-9485]{Kosuke Namekata}
\affil{The Hakubi Center for Advanced Research, Kyoto University, Yoshida-Honmachi, Sakyo, Kyoto 606-8501, Japan}
\affil{Department of Physics, Kyoto University, Kitashirakawa-Oiwake-cho, Sakyo, Kyoto 606-8502, Japan}
\affil{\nasa}
\affil{Department of Physics, The Catholic University of America, 620 Michigan Ave NE, Washington, DC 20064, USA}
\affil{Division of Science, National Astronomical Observatory of Japan, NINS, 2-21-1 Osawa, Mitaka, Tokyo 181-8588, Japan}
\email[]{}

\author[0000-0001-7624-9222]{Pascal Petit}
\affil{Institut de Recherche en Astrophysique et Plan\'{e}tologie, Universit\'{e} de Toulouse, CNRS, CNES, 14 avenue \'{E}douard Belin, 31400 Toulouse, France}
\email[]{}

\author[0000-0003-4452-0588]{Vladimir S. Airapetian}
\affil{\nasa}
\affiliation{Department of Physics, American University, 4400 Massachusetts Ave NW, Washington, DC 20016, USA}
\email[]{}

\author[0000-0003-0332-0811]{Hiroyuki Maehara}
\affil{Okayama Branch Office, Subaru Telescope, National Astronomical Observatory of Japan, NINS, 3037-5 Honjo, Kamogata, Asakuchi, Okayama 719-0232, Japan}
\affil{Astronomical Observatory, Kyoto University, Kitashirakawa-Oiwake-cho, Sakyo, Kyoto 606-8502, Japan}
\email[]{}

\author[0000-0002-0412-0849]{Yuta Notsu}
\affil{Laboratory for Atmospheric and Space Physics, University of Colorado Boulder, 3665 Discovery Drive, Boulder, CO 80303, USA}
\affil{National Solar Observatory, 3665 Discovery Drive, Boulder, CO 80303, USA}
\email[]{}

\author[0000-0001-5371-2675]{Aline A. Vidotto}
\affil{Leiden Observatory, Leiden University, P.O. Box 9513, 2300 RA Leiden, The Netherlands}
\email[]{}

\author[0000-0001-7115-2819]{Keith Gendreau}
\affil{\nasa}
\email[]{}

\author[0000-0003-2490-4779]{Sandra V. Jeffers}
\affil{Max Planck Institute for Solar System Research, Justus-von-Liebig-weg 3, 37077 G\"ottingen, Germany}
\email[]{}

\author[0000-0001-5522-8887]{Stephen Marsden}
\affil{Centre for Astrophysics, University of Southern Queensland Toowoomba, 487-535 West Street, 
Toowoomba, Queensland 4350, Australia}
\email[]{}

\author[0000-0002-4996-6901]{Julien Morin}
\affil{LUPM, Universit\'e de Montpellier, CNRS, Place Eug\`ene Bataillon, F-34095 Montpellier, France}
\email[]{}

\author[0000-0003-1978-9809]{Coralie Neiner}
\affil{LIRA, Paris Observatory, PSL University, CNRS, Sorbonne University, Universit\'e Paris Cit\'e, CY Cergy Paris University, 5 place Jules Janssen, 92195 Meudon, France}
\email[]{}

\author[0000-0002-8090-3570]{Rishi R. Paudel}
\affil{\nasa}
\affil{University of Maryland, Baltimore County, 1000 Hilltop Circle, Baltimore, MD 21250, USA}
\affil{Space Research Centre, Faculty of Technology, Nepal Academy of Science and Technology, Khumaltar, Lalitpur, Nepal}
\email[]{}

\author[0000-0001-9588-1872]{Daisaku Nogami}
\affil{Department of Astronomy, Kyoto University, Kitashirakawa-Oiwake-cho, Sakyo, Kyoto 606-8502, Japan}
\email[]{}

\author[0000-0003-1206-7889]{Kazunari Shibata}
\affil{Kwasan Observatory, Kyoto University, 17 Ohmine-cho, Kita-Kazan, Yamashina, Kyoto 607-8471, Japan}
\affil{Department of Environmental Systems Science, Doshisha University, 1-3 Tataramiyakodani, Kyotanabe, Kyoto 610-0394, Japan}
\email[]{}

\begin{abstract}
Recent simultaneous multiwavelength observations of a nearby young solar-type star EK Dra in the optical, H$\alpha$ spectrum, and X-ray, have provided evidence for stellar prominence eruptions associated with superflares. The large prominence eruption is suggested to have been caused by a large mid-latitude spot on the polarity inversion lines near the stellar limb from the concurrent Zeeman Doppler Imaging (ZDI) and optical photometry by the TESS.
In this study, we perform starspot mapping for the TESS data of EK Dra to investigate the relation of starspots and magnetic fields from the photometry and ZDI.
We also explore the multiwavelength rotational variability ascribed to starspots and active regions for the TESS, B-band, H$\alpha$, and X-ray light curves.
As a result, we find that (i) spot locations deduced from the TESS light curve are mostly consistent with the intensity map from the ZDI except for a polar spot, and (ii) the H$\alpha$ light curve exhibits clear periodicity with respect to the TESS light curve because the H$\alpha$ line is radiated around spots in the chromosphere. The X-ray light curve does not show such association probably because of multiple spots on high activity level and extended spatial structure of coronal active regions.
The results provide clues to explore their association with stellar flares at different heights of active regions in chromospheric and coronal lines. Our study also enables us to quantify the stellar XUV radiation from the magnetic fields of active stars toward understanding atmospheric evolution of exoplanets. 
\end{abstract}

\keywords{\uat{Starspots}{1572} --- \uat{G dwarf stars}{556} --- \uat{Solar analogs}{1941} --- \uat{Flare stars}{540} --- \uat{Stellar coronae}{305} --- \uat{Stellar atmospheres}{1584} --- \uat{Stellar magnetic fields}{1610} --- \uat{Doppler imaging}{400} --- \uat{Zeeman-Doppler imaging}{1837} --- \uat{Astrostatistics}{1882}}

\section{Introduction}\label{sec:intro}
Solar and stellar flares are intense explosions in the solar and stellar atmospheres by releasing magnetic energy around sunspots and starspots, and they have been observed from radio to X-rays \citep[for reviews,][]{Benz10, Shibata11,Benz17, Toriumi19, Kowalski24}.
Starspots have been extensively studied through ground-based observations of photometry and spectroscopy for active young stars, cool stars (G-, K-, and M-dwarfs), and RS CVn-type stars \citep[for reviews,][]{Berdyugina05,Strassmeier09}, and interferometry for active giant stars \citep[e.g.,][]{Roettenbacher16}. 
In particular, the Kepler Space Telescope \citep{Koch10} and the Transiting Exoplanet Survey Satellite \citep[TESS;][]{Ricker15} have facilitated to investigate starspots \citep[e.g.,][]{Basri10,Basri13} and stellar flares \citep[e.g.,][]{Hawley14,Davenport16} with their high precision and long-term photometry.
Many superflares have been reported on a large number of solar-type stars \citep{Maehara12, Shibayama13, Maehara15, Notsu19, Okamoto21, Guenther20, Feinstein20, Tu21, Jackman21, Namekata22, Namekata22b, Pietras22, Crowley22, Howard22, Namekata23, Feinstein24,Vasi24,Namekata25,Namekata25b} in close connection with large spots and magnetic fields on the stellar surface \citep{Shibata13, Notsu13,Notsu15b,Maehara17, Namekata19, kepler17, Yamashita22,Namekata24,Tokuno25}. 
In particular, young solar-type stars have been suggested to possess strong magnetic fields \citep{Marsden06,Waite11, Vidotto14, Folsom16, Folsom18, Kochukhov20,Yamashita25} and cause frequent superflares \citep{Audard99, Audard00, Namekata22b, Yamashita22, Colombo22, Feinstein24}.
\cite{Namekata22} has reported a large filament eruption associated with a superflare on a young solar-type star EK Draconis \citep[EK Dra, spectrum type of G1.5V;][]{Strassmeier98}. The dynamics of the eruptive filament is explained by utilizing a Sun-as-a-star analysis through spatially integrated spectrum of solar eruptive events \citep[e.g.,][]{Namekata_sun,Otsu22,Otsu24,Otsu24b,Otsu25} and performing a hydrodynamic simulation of an eruptive filament in an expanding magnetic loop \citep[][]{Shibata24}.

Recent multiwavelength observations of EK Dra simultaneously by the TESS, H$\alpha$ spectrum, and X-ray, have provided evidence of large prominence eruptions associated with superflares on a solar-type star \cite[][hereafter referred to as \citetalias{Namekata23}]{Namekata23}.
Specifically, two prominence eruptions were observed in ${\rm H}\alpha$ emission with the blueshifted velocities of 690 km s$^{-1}$ and 430 km s$^{-1}$. 
The larger event was additionally accompanied by possible X-ray dimming interpreted as a signature of a stellar coronal mass ejection (CME).
To examine the relation between prominence eruptions and the distribution of magnetic fields on the stellar surface \citep[][hereafter referred to as \citetalias{Namekata24}]{Namekata24}, we conducted data-driven modeling combined with Zeeman Doppler Imaging (ZDI).
The semi-analytical free-fall model indicates that the prominence eruption could have occurred near the stellar limb within 12$^\circ$-16$^\circ$ and ejected at an angle of 15$^\circ$-24$^\circ$ relative to the line of sight. 
The hydrodynamic simulation also reproduces the dynamics of a falling prominence in an expanding magnetic loop and supports a scenario that the magnetic loop subsequently evolved into stellar CMEs in consistent with the X-ray dimming. 
The concurrent ZDI suggests that the superflare was likely caused around spots on the polarity inversion line (PIL) close to the stellar limb or alternatively a polar spot with a single
polarity.
In addition, the rotational phase of the TESS light curve implies the presence of spots near the limb \citepalias{Namekata23}, but it is unknown where the spots are actually present on the surface.
Therefore, it is important to investigate the relation between the spot locations and occurrence of a superflare by mapping starspots on the stellar surface \citep[e.g.,][]{Ikuta20,Bicz22,Ikuta23,Bicz24a,Bicz24b}, especially for the quasi-periodic and temporally variable light curve of G-dwarfs. In particular, there is also suggested to be many spot groups in multiple active regions for G-dwarfs \citep{Schrijver20, Takasao20, kepler17}, and there is much information about starspot properties in the temporal photometric variability \citep[e.g.,][]{Walkowicz13,Basri18,Basri20}.
Thus, based on this study (Paper I\hspace{-.1em}I\hspace{-.1em}I), we conduct starspot mapping for the TESS light curve with three spots, and the deduced spot distributions indicate that the superflare is likely to be caused around a spot near the limb in agreement with the ZDI \citepalias{Namekata24}.

In \citetalias{Namekata23} and \citetalias{Namekata24}, we also have a unique opportunity to investigate the relation between starspots and coronal properties from simultaneous multiwavelength observations of a young solar-type star EK Dra.
Starspots and active regions are also observed in the H$\alpha$ spectrum and X-ray in association with the stellar chromosphere and corona. 
The rotational variation of the H$\alpha$ spectrum has been reported to be ascribed to starspots for M-dwarfs \citep{Namekata20,Maehara21, Schoefer22,Odert25, Notsu25} and solar-type stars \citep{Namekata22b} as a byproduct of spectroscopic monitoring observations of stellar flares \citep[e.g., ][]{Honda18, Notsu24,Kajikiya25,Kajikiya25b,Ichihara25}.
Starspots are also closely connected with X-ray radiations from the stellar corona through the dissipation of magnetic fields \citep[e.g.,][]{Guedel97,Notsu17, Takasao20,Shoda21,Washinoue22,Washinoue23,Shoda24}.
Thus, the chromospheric and coronal heating can be constrained by exploring the light curves of solar-type stars as represented by EK Dra in the optical, H$\alpha$ spectrum, and X-ray \citep{Toriumi20, Airapetian21, Toriumi22a, Toriumi22b,Namekata_XUV}.

EK Dra is a nearby active G-dwarf with the age of 30 to 125 Myr in a binary system with the orbital period of $45 \pm 5$ yr \citep{Konig05,Waite17,Senavci21} (Table \ref{tb:stellar}). 
EK Dra exhibits a high level of stellar magnetic activity with a hot and dense corona \citep{Guedel95}, starspots from ground-based photometry \citep{Chugainov91, Scheible94,Dorren94, Lockwood97, Strassmeier97, Strassmeier98,Froehlich02, Messina03, Jarvinen05,Berdyugina05b}, 
starspots and magnetic fields through the Doppler Imaging \citep[DI;][]{Strassmeier98,Jarvinen18,Senavci21,Gorgei25} and ZDI  \citep[][\citetalias{Namekata24}]{Rosen16, Jarvinen07, Waite17}, and frequent superflares \citep{Audard99,Audard00, Namekata22,Namekata23}.

In this study (Paper I\hspace{-.1em}I\hspace{-.1em}I), we conduct starspot mapping for the TESS light curve of a young solar-type star EK Dra, for the purpose of investigating (i) their correspondence to the starspots and magnetic fields from the ZDI in \citetalias{Namekata24} and (ii) the comparison of the spot locations from the TESS light curve with the H$\alpha$ and X-ray light curves in correspondence to the chromosphere and corona. We also discuss (iii) the relation between spot properties and the occurrence of superflares reported in \citetalias{Namekata23}.
The remainder of this paper is organized as follows. In Section \ref{sec:dataset}, we briefly introduce the data of the TESS, ground-based B-band photometry, H$\alpha$ spectrum, X-ray, and the result of the ZDI. In Section \ref{sec:method}, we introduce the numerical setup and periodic analysis for multiwavelength variability. In Section \ref{sec:result}, we discuss the deduced result in terms of comparison with the ZDI, multiwavelength variability especially in the H$\alpha$ and X-ray, and relation between spot properties and superflares. In Section \ref{sec:conclusion}, we conclude this paper and future prospects. In Appendix \ref{sec:appendix}, we show the result of spot mapping for three TESS sectors. In Appendix \ref{sec:t_em},
we describe the solar and stellar coronal properties.

\begin{deluxetable*}{lc}[tbhp!]
\tablecaption{Stellar parameters \label{tb:stellar}}
\tabletypesize{\footnotesize}
\tablehead{
\colhead{Stellar parameter} & \colhead{EK Dra} }
\startdata
Effective temperature $T_{{\rm eff}}$ (K)&$5700 \pm 70$ $^{\rm d}$ \\
Rotation period $P_{{\rm rot}}$ (day)&  $2.766 \pm 0.002$ $^{\rm e}$ \\
Stellar radius $R_{\rm star}$ ($R_{\rm Sun}$)  &$0.94 \pm 0.07$ $^{\rm e}$ \\
Stellar mass $M_{\rm star}$ ($M_{\rm Sun}$)  &$0.95 \pm 0.04$ $^{\rm e}$ \\
Surface gravity $\log g$& $4.47 \pm 0.08$ \\
Inclination angle $i$ (deg)&$60\pm5$ $^{\rm e}$ \\ 
Equatorial rotational velocity $\Omega_{\rm eq}$ (rad day$^{-1}$) \tablenotemark{\rm a} & $2.50 \pm 0.08$ $^{\rm e}$ \\ 
Rotational shear $\Delta \Omega$ (rad day$^{-1}$) \tablenotemark{\rm a}  &  $0.27^{+0.24}_{-0.26}$ $^{\rm e}$ \\ \hline
Spot temperature $T_{\rm spot}$ (K) \tablenotemark{\rm b}& $4069 \pm 28$ \\ 
Spot relative intensity  $f_{\rm spot}$ in TESS- and B-bands \tablenotemark{\rm b}& $0.27 \pm 0.01$, $0.10 \pm 0.01$ \\ 
\hline
Limb-darkening coefficients ($c_1$, $c_2$, $c_3$, $c_4$) in TESS- and B-bands \tablenotemark{\rm c} & ($0.86$, $-0.96$, $1.35$, $-0.54$), ($0.56$, $-0.64$, $1.63$, $-0.68$) \\ 
\enddata
\tablenotetext{\rm a}{
The angular velocity at each latitude $\Phi$ is represented by solar differential rotation as $\Omega(\Phi) = \Omega_{\rm eq} - \Delta \Omega \sin^2 \Phi$ \citep[][]{Waite17}. The parameters of the spotted flux in the model correspond to $P_{\rm eq} = 2 \pi/ \Omega_{\rm eq}$ and $\kappa = \Delta \Omega/  \Omega_{\rm eq}$ (Equation \ref{eq:rot}).
}
\tablenotetext{\rm b}{
The spot temperature is formulated by the stellar effective temperature $T_{{\rm eff}}$ with Equation \ref{eq:spottemp}, and the spot relative intensities $f_{\rm spot}$ for TESS-band and B-band \citep{bessel90} are calculated by Equation \ref{eq:relspot}.
}
\tablenotetext{\rm c}{
The stellar limb-darkening coefficients for TESS-band and B-band are characterized by the effective temperature $T_{\rm eff}$ and surface gravity $\log g$ under the solar metallicity \citep[][]{Claret23}. {We adopt the coefficient values of ($T_{\rm eff}$, $\log g$)=($5700$, $4.5$), which are derived by their Method 1.}
}
\tablerefs{}{$^{\rm d}$ \cite{Konig05}; $^{\rm e}$ \cite{Waite17};

}
\end{deluxetable*}

\begin{figure*}[thbp!]
\plotone{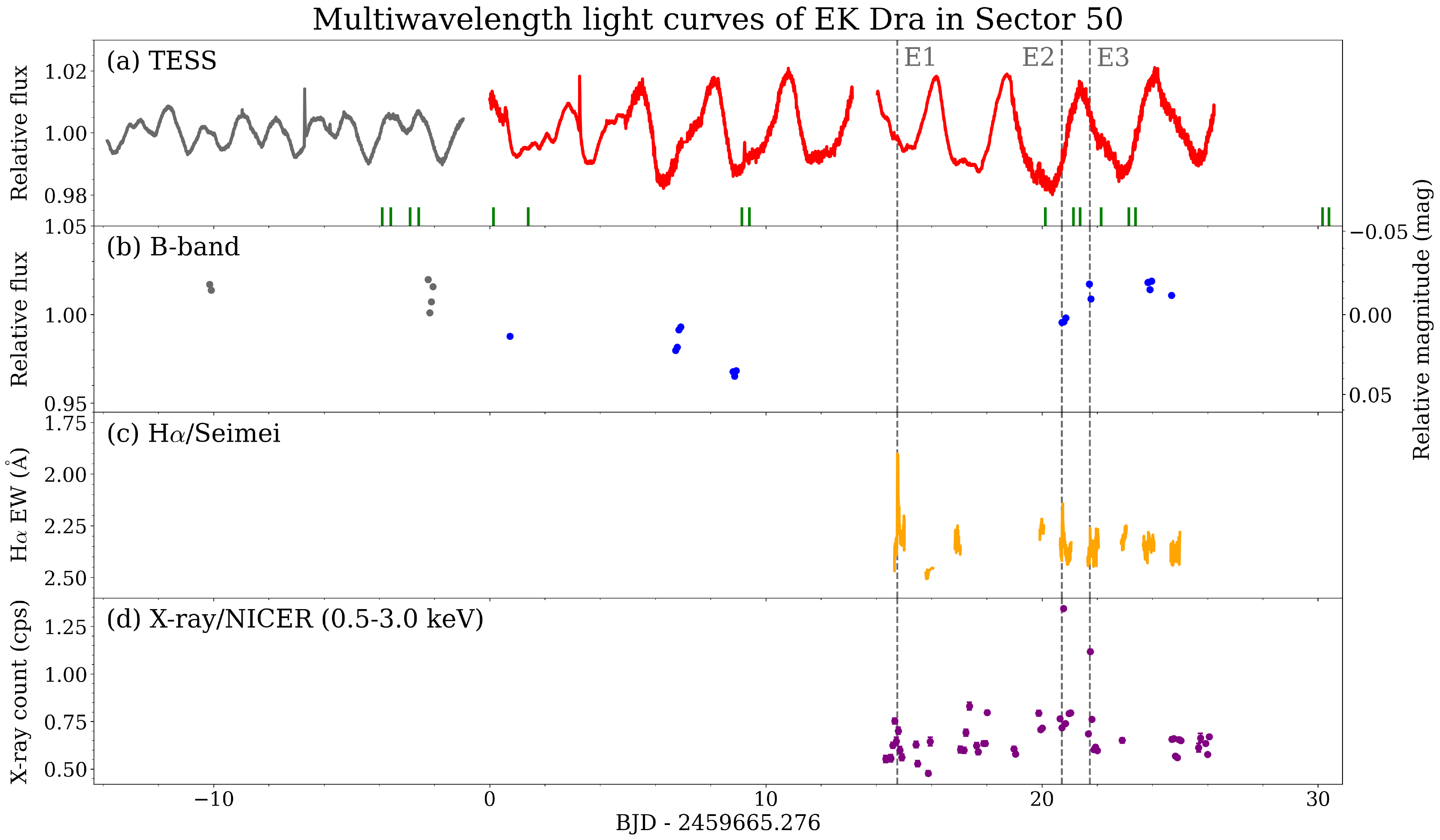} 
\caption{Multiwavelength light curves of the (a) TESS, (b) B-band, (c) H$\alpha$, and (d) X-ray, in the Sector 50 (red, blue, orange, and purple). 
We also exhibit the TESS and B-band light curves in the second half of the Sector 49 (gray).
The vertical dashed lines in all the panels (gray) represent the time of the prominence eruptions E1, E2, and E3 \citepalias[][]{Namekata23}.
In the panel (a), the vertical lines (green) represent the time of obtaining the spectra with the TBL/Neo-NARVAL for the ZDI \citepalias[Table 1 in][]{Namekata24}.
}
 \label{fig:lightcurve}
\end{figure*}

\section{Observations} \label{sec:dataset}
In \citetalias[][]{Namekata23}, we performed multiwavelength monitoring observations of EK Dra by the TESS, H$\alpha$ spectrum, and X-ray to investigate the properties of flares and plasma eruptions.
In \citetalias[][]{Namekata24}, we conducted the concurrent spectropolarimetric observations of EK Dra to investigate the relation between plasma eruptions and magnetic fields through the Zeeman Doppler Imaging (ZDI).

In this study (Paper I\hspace{-.1em}I\hspace{-.1em}I), we utilize the multiwavelength and spectropolarimetric data to compare the spot map deduced from the TESS light curve with the magnetic field from the ZDI and to explore multiwavelength variability ascribed to spots and active regions. Then, we additionally incorporate a ground-based B-band light curve.
As shown in Figure \ref{fig:lightcurve}, we briefly describe multiwavelength data for the TESS (Section \ref{sec:tess}), ground-based B-band photometry (Section \ref{sec:ground}), H$\alpha$ spectrum with low-resolution spectroscopy (Section \ref{sec:kools}), and X-ray (Section \ref{sec:nicer}).
We also describe the observations of high-resolution spectropolarimetry and results of the ZDI (Section \ref{sec:obs}).

\subsection{Optical photometry -- TESS} \label{sec:tess}
EK Dra (TIC159613900) was observed with ten-minute cadence by the TESS in its Sector 50 (March 26 to April 22 in 2022) through the TESS extended mission (Cycle 4).
We retrieve the Full Frame Images (FFI) from the MAST (Mikulski Archive for Space Telescopes) Portal website using \texttt{eleanor} \citep{Feinstein19}.
We calibrate the light curve with the principal component analysis as the PCA-flux.
The light curve exhibits a quasi-periodic modulation ascribed to spots that evolve in the size and location along with the stellar differential rotation.
In addition, we extract outliers including flares from an interpolated flux using \texttt{stella} \citep{Feinstein20} for spot mapping (Section \ref{sec:spotmap}). 
The \texttt{stella} code detects flares from the TESS data with the two-minute cadence using neural networks, and we adopt their pretrained model with the detected flares \citep{Guenther20}.
We reproduce the interpolated flux with two-minute cadence using Gaussian process with modified quasi-periodic kernel \citep{celerite} in \texttt{tinygp} \citep{tinygp} because there is only FFI data with the ten-minute cadence for the Sector 50, and \texttt{stella} is only adopted to the TESS data with the two-min cadence.
We use the data in the Sector 50 for simplicity, and we also conduct spot mapping for the data in its Sectors 48 and 49 (January 28 to March 26 in 2022)
together with those in the Sector 50 (Appendix \ref{sec:appendix}).

\begin{deluxetable*}{lcc}[tbhp!]
\tablecaption{Previous Doppler Imaging$^{\rm a}$ and Zeeman Doppler Imaging$^{\rm b}$ of EK Dra \label{tb:doppler}}
\tabletypesize{\scriptsize}
\tablehead{
\colhead{Year} & \colhead{Spot latitude and temperature difference} & \colhead{Reference} 
}
\startdata
1995/3 & low to mid ($\sim 40^{\circ}$) and high (70 to 80$^{\circ}$), 400 - 1200 K & \cite{Strassmeier98}$^{\rm a}$ \\  
2001/7-8  &equator to high ($\sim 70^{\circ}$), 500 K& \cite{Jarvinen07}$^{\rm a}$ \\
2002/2-3,8   &equator to intermediate ($<60^{\circ}$), 500 K& \cite{Jarvinen07}$^{\rm a}$ \\ 
2007/7 & high ($>80^{\circ}$)  & \cite{Jarvinen09}$^{\rm a}$  \\ 
2007/1, 2012/1  &  equator to intermediate ($<60^{\circ}$)  & \cite{Rosen16}$^{\rm b}$  \\ 
2006/12, 2007/1-2, 2008/1 & low to mid and intermediate ($40^{\circ}$ to $70^{\circ}$)    & \cite{Waite17}$^{\rm b}$  \\ 
2012/1 & low to mid and pole    & \cite{Waite17}$^{\rm b}$  \\
2015/4  & low ($3^{\circ}$ and $8^{\circ}$) and mid ($43^{\circ}$ and $48^{\circ}$), 280 - 990 K  & \cite{Jarvinen18}$^{\rm a}$\\ 
2015/7  & equator to mid ($< 50^{\circ}$) and pole ($77^{\circ}$)   & \cite{Senavci21}$^{\rm a}$   \\  
2021/2, 2023/4-6,10, 2024/3-4,7 & equator to high-latitude in both hemisphere & \cite{Gorgei25}$^{\rm a}$ \\ \hline
2022/3-4 & 20-40$^{\circ}$ and pole    & \citetalias{Namekata24} (Section \ref{sec:obs})$^{\rm b}$  \\
\enddata
\end{deluxetable*}

\subsection{Optical photometry -- Ground-based B-band} \label{sec:ground}
We performed photometric monitoring observations of EK Dra in B-band with the 11.5cm telescope at Okayama in the Sector 49 and 50 (March 16 to April 20 in 2022). 
We calibrate a relative photometry with the comparison star TYC 4176-314-1
with the distance of $\sim 8.1'$ from EK Dra.
We derive the relative flux by normalizing the photometry with its average in the observation period.

\subsection{Optical spectroscopy -- Seimei/KOOLS-IFU} \label{sec:kools}
We conducted spectroscopic monitoring observations of EK Dra by the Kyoto Okayama Optical Low-dispersion Spectrograph with optical-fiber Integral Field Unit \citep[KOOLS-IFU;][]{Matsubayashi19,Matsubayashi25} on the 3.8m Seimei telescope at Okayama observatory of Kyoto University \citep{Kurita20} from April 10 to 20 in 2022.
The KOOLS-IFU is an optical spectrograph with a spectral resolution of $R\sim2000$ in the wavelength range from 5800 to 8000 ${\rm \AA}$.
The exposure time was set to be either 60 or 120 seconds, depending on weather conditions.
The data reduction are performed in the procedure of \citetalias{Namekata23}, and the one-dimensional spectrum is produced by calibrating the wavelength and by normalizing with the continuum for each time.
We compute the equivalent width (EW) of the H$\alpha$ emission centered on 6562.8${\rm \AA}$ by integrating its emission from 6562.8-10${\rm \AA}$ to 6562.8+10${\rm \AA}$ after normalized by the continuum flux.
We note that the EW is defined to be negative in terms of the emission.

\subsection{X-ray spectroscopy -- NICER} \label{sec:nicer}
We conducted spectroscopic observations by the Neutron star Interior Composition Explorer \citep[NICER;][]{NICER} on the International Space Station (ISS) from April 10 to 21 in 2022. 
The exposure time is set to be a few minutes to a few tens of minutes, depending on the dates.
We retrieve the data from NASA HEASARC webpage and reproduce the X-ray Timing Instrument (XTI).
The XTI has a large collection area between 0.2 and 12 keV, and we produce the X-ray light curve in the energy range from 0.5 to 3 keV, where the contamination from the background is relatively negligible.

\subsection{Zeeman Doppler Imaging -- TBL/Neo-NARVAL} \label{sec:obs}
We conducted spectropolarimetric observations with Neo-NARVAL \citep{narval} mounted on the T\'{e}lescope Bernard Lyot (TBL) at the observatoire du Pic du Midi from March 22 to April 26 in 2022. 
We obtained 16 sequences with enough signal-to-noise ratio (S/N) in the wavelength range from 380 to 1,050 nm with the wavelength resolution of $R\sim65000$, and we restrict the analysis of the data redder than 470 nm to get rid of the bluest fraction of spectra affected by a very low S/N.
The line profiles are produced with the procedure of the Least Squares Deconvolution \citep[LSD;][]{Donati97,Kochukhov10}, which extracts the average line profiles in Stokes $I$ and $V$ from each spectra.
The intensity map and large-scale magnetic fields are reconstructed from the LSD profiles respectively in Stokes $I$ and $V$ with the Zeeman Doppler Imaging (ZDI) technique. 
The ZDI is performed using the values of the inclination angle $i = 60^{\circ}$, $v \sin i$ = 16.4 km s$^{-1}$ \citep{Waite17}, reference time of 2661.39 (BJD-2457000), rotation period $P_{\rm rot}=$ 2.766 day, and rigid rotation.
We note that a search for a solar-like differential rotation is inconclusive from the method of the sheared image \citep{Petit02}.

The intensity map from Stokes $I$ shows that the visible pole is occupied by the dark spot, and large dark spots are latitudinally extended around 20 to 40 deg.
The brightest features are concentrated on a latitude of 60 deg, and bright and dark features are distributed around low latitudes.
These features are resolved with the spatial resolution estimated to be $\sim$ 25 deg from the wavelength resolution and phase coverage \citepalias[Appendix B in][]{Namekata24}.
The previous spot properties from the DI and ZDI are summarized in Table \ref{tb:doppler}, and the spots on high latitudes are consistent with the previous studies \citep{Strassmeier98, Waite17, Senavci21} except for \cite{Jarvinen18}, which does not reconstruct a polar spot.
In particular, the very symmetric configuration is 
close to the maps in \cite{Strassmeier98} and \cite{Senavci21}, although non-axisymmetric spots on high latitude are reported with respect to the pole.
The radial magnetic field from Stokes $V$ shows that 
there is strong negative radial field at high latitude and prominent negative toroidal field. 
The polar spot from the intensity map is consistent with the strong radial magnetic field close to the pole in larger extension to the latitude of 30 deg. 
The general magnetic fields are in overall agreement with the maps of magnetic fields in \cite{Waite17}.

\section{Method} \label{sec:method}

\subsection{Spot mapping for the TESS data} \label{sec:spotmap}

Table \ref{tb:stellar} lists stellar parameters, derived spot temperature and intensity, and limb-darkening coefficients in TESS- and B-bands for EK Dra.
The spot temperature $T_{\rm spot}$ is derived from the quadratic formula of the stellar effective temperature  $T_{\rm eff}$ \citep{Herbst21}:
\begin{equation} \label{eq:spottemp}
T_{\rm spot} = -3.58 \times 10^{-5} T_{\rm eff}^2 + 0.801 T_{\rm eff}+666.5 .
\end{equation}

Thereby, the spot relative intensity $f_{\rm spot}$ in TESS- and B-bands is calculated by the stellar effective temperature $T_{\rm eff}$ and spot temperature $T_{\rm spot}$ \citep[Equation 3 in][]{Ikuta23}:
\begin{equation} \label{eq:relspot} 
f_{\rm spot} = \frac{\int R(\lambda) B(\lambda,T_{\rm spot}) d\lambda}{\int R(\lambda) B(\lambda,T_{\rm eff}) d\lambda},
\end{equation}
where $R(\lambda)$ and $B(\lambda,T)$ are the response function for each of the band and the Plank function with each wavelength $\lambda$, respectively\footnote{For simplicity, we approximate the spot spectrum using the Plank function, although the Plank function and stellar spectrum have intrinsic uncertainties to reproduce the spot spectrum \citep{Smitha25}.}.
The stellar limb-darkening law is adopted as the following formula for the cosine of the angle to the line of sight $\mu$: 
\begin{equation} \label{eq:limb}
I(\mu)/I(1)= 1 - \sum_{n=1}^4 c_n(1-\mu^{n/2}).
\end{equation}
The coefficients $c_k$ are specified by the stellar effective temperature and surface gravity and derived from the nonlinear limb-darkening law in the TESS- and B-bands \citep{Claret23}.

We implement a code for starspot mapping \citep{Ikuta20} using an adaptive parallel tempering \citep[e.g.,][]{Vousden16} with an importance sampling algorithm \citep{Kass95, Diaz14} to deduce parameters and compare models with different number of spots. We apply the code to the TESS light curves of M-dwarf flare stars \citep{Ikuta23,Mori24}.
The code is adopted with an analytical modeled flux with a specified number of circular spots \citep[\texttt{macula};][]{kipping12}\footnote{The code of the flux model in JAX is available on Zenodo (\url{https://doi.org/10.5281/zenodo.18232159}), as developed on GitHub 
(\url{https://github.com/KaiIkuta/StarspotMapping}).}, and the model is specified in the stellar and spot parameters: the stellar inclination angle $i$ (deg); equatorial rotation period $P_{\rm eq}$ (day); degree of differential rotation $\kappa$; spot relative intensity $f_{\rm spot}$; latitude $\Phi_k$ (deg); initial longitude $\Lambda_k$ (deg); reference time  $t_{{\rm ref},k}$ (day) (the time at the midpoint of the interval over
which the spot has its maximum radius); maximum radius $\alpha_{{\rm max},k}$ (deg); emergence duration ${\cal I}_k$ (day); decay duration ${\cal E}_k$ (day);
stable duration ${\cal L}_k$ (day).
The rotation period at the latitude $\Phi$ under a differential rotation \citep[Equation 16 in][]{Ikuta20} is represented by
\begin{equation} \label{eq:rot}
P(\Phi)=\frac{P_{\rm eq}}{1-\kappa \sin^2 \Phi}.
\end{equation}
We fix the inclination angle $i$ and spot relative intensity $f_{\rm spot}$ with the literature values (Table \ref{tb:stellar}) due to the parameter degeneracies, as described in \cite{Ikuta23}.
In this study, we adopt temporally variable spot radius $\alpha(t)$ \citep[][]{kepler17,Ikuta20} as the spot radius increase and decrease linearly with time by setting ${\cal L}_k=0$ \citep[Equation 14 in][]{Ikuta20}:

\begin{eqnarray} \label{eq:spotrad}
  \alpha_k (t)  = \left\{ \begin{array}{l}
    \alpha_{{\rm max},k} \{t-(t_{{\rm ref},k}-{\cal I}_k)\}/{\cal I}_k \\ \hspace{15mm} (t_{{\rm ref},k}-{\cal I}_k< t \leq t_{{\rm ref},k}) \\
    \alpha_{{\rm max},k}\{(t_{{\rm ref},k}+{\cal E}_k)-t\}/{\cal E}_k \\ \hspace{15mm} (t_{{\rm ref},k}<t \leq t_{{\rm ref},k}+{\cal E}_k) \\
    0  \hspace{13mm}  (t \leq t_{{\rm ref},k}-{\cal I}_k, t_{{\rm ref},k}+{\cal E}_k<t ) 
  \end{array} \right.
\end{eqnarray}
We note that we adopt three more parameters for each spot than those in \cite{Ikuta23}: for M-dwarfs, the spot radius can be assumed to be stable because the amplitudes of the light curves are approximately constant in the observation period of a TESS Sector \citep[e.g.,][]{Davenport20}.
We set the time $t$ and spot reference time $t_{\rm ref}$ from the time of 2665.276 (BJD -2457000) (the start of the Sector 50).
We also set log-normal and normal prior distributions for the equatorial period $P_{\rm eq}$ and degree of differential rotation $\kappa$ from the literature values (Table \ref{tb:stellar}) by calculating 
$P_{\rm eq} = 2 \pi/ \Omega_{\rm eq}$ and $\kappa = \Delta \Omega/  \Omega_{\rm eq}$ from the solar differential rotation as $\Omega(\Phi) = 2 \pi / P(\Phi) = \Omega_{\rm eq} - \Delta \Omega \sin^2 \Phi $ \citep[Equation 5 in][]{Waite17}.
We also set log-uniform prior distributions for emergence and decay durations ${\cal I}$ and ${\cal E}$, and uniform prior distributions are set for other parameters. Each of the spot is discerned by the range of latitude $\Phi$ (Table \ref{tb:para}). 
We optimize the light curves by the two- and three-spot models to reproduce overall features ascribed to large spot groups. This is because the residual only decreases if more than three spots are adopted \citep{Ikuta23}.
We note that more spots are suggested to be on the surface than the number of local minima per one rotational period \citep{kepler17}, and there are significant degeneracies between the number of spots and spot sizes because the unspotted level may not be determined only from a monochromatic light curve \citep{Basri18, Ikuta20,Luger21a, Luger21b}. In this study, we do not take into account bright spots (faculae) because the contributions of spots and faculae would cancel out in the monochromatic light curve \citep{Luger21b}.
In addition, faculae are expected to contribute only weakly to the variability of the light curve since EK Dra lies in the regime of spot-dominated stars \citep{Montet17} and faculae are thought to be distributed broadly across the entire surface \citep{Johnson21}.

\begin{deluxetable*}{lccc}[tbhp!]
\tablecaption{EK Dra (Sector 50)
\label{tb:para}}
\tabletypesize{\scriptsize}
\tablehead{
\colhead{Deduced Parameters} &  \colhead{Two-spot Model} & \colhead{Three-spot Model} & \colhead{Prior Distribution\tablenotemark{{\rm a}}}  
}
\startdata
(Stellar parameters) &&& \\
1.  Equatorial period $P_{{\scriptsize \textrm{eq}}}$ (day) &$2.6408 ^{+ 0.0001}_{- 0.0001}$ &$ 2.6181 ^{+ 0.0001}_{-0.0003}$&${\cal N}_{\log} (2.51,0.08^2)$\tablenotemark{{\rm b}}\\
2.   Degree of differential rotation $\kappa$ &$ 0.0000^{+ 0.0001}_{- 0.0000}$ &$ 0.0748 ^{+ 0.0002}_{- 0.0002}$ &${\cal N} (0.108,0.104^2)$\tablenotemark{{\rm b}}\\
(Spot parameters) &  & &   \\
(First spot) &  &  &  \\
3.     Latitude $\Phi_1$  (deg) & $ 50.92 ^{+ 0.05 }_{- 0.05}$&$ 25.06 ^{+ 0.08}_{- 0.02}$ &{${\cal U} (\Phi_2,\Phi_3)$\tablenotemark{{\rm c}}}\\
4.     Initial longitude $\Lambda_1$ (deg) &$ 110.46 ^{+ 0.04 }_{- 0.04}$ &$ 101.22 ^{+ 0.05 }_{- 0.04}$ &${\cal U}  (-180.00,180.00)$\\
5.     Reference time $t_{1}$ (day) & $ 21.215 ^{+ 0.006 }_{- 0.006 }$ &$ 20.876 ^{+ 0.009 }_{- 0.009}$&${\cal U} (0.00,26.229)$\\
6.     Maximum radius $\alpha_{{ \textrm{max,1}}}$ (deg)&$ 11.332 ^{+ 0.004 }_{- 0.004 }$ &$ 9.732 ^{+ 0.004}_{- 0.003}$ &${\cal U}(0.01,20.00)$\\
7.     Emergence duration ${\cal I}_1$ (day)&$ 59.493 ^{+ 0.046 }_{- 0.056}$ &$ 78.017 ^{+ 0.126}_{- 0.156}$
&${\cal U}_{\log} (0.000,200.000)$\\
8.     Decay duration ${\cal E}_1$ (day)&$ 32.014 ^{+ 0.056}_{- 0.115}$&$ 29.062 ^{+ 0.108 }_{- 0.095 }$
 &${\cal U}_{\log} (0.000,200.000)$\\
(Second spot)&  &  &   \\
9.     Latitude $\Phi_2$  (deg) &$ 2.02 ^{+ 0.07 }_{- 0.06}$& $ 25.05^{+ 0.09}_{- 0.02}$&{${\cal U} (-90.00,\Phi_1)$\tablenotemark{{\rm c}}}\\
10.     Initial longitude $\Lambda_2$  (deg)&$ -129.04 ^{+ 0.04}_{- 0.03}$&$ -134.76 ^{+ 0.02}_{- 0.04}$&${\cal U}  (-180.00,180.00)$\\
11.     Reference time $t_{2}$ (day)  &$ 6.089 ^{+ 0.001 }_{- 0.001}$ &$ 7.752 ^{+ 0.004 }_{- 0.004}$&${\cal U} (0.00,26.229)$ \\
12.     Maximum radius $\alpha_{{\textrm{max,2}}}$  (deg)&$ 11.053 ^{+ 0.005 }_{- 0.005}$&$ 11.144 ^{+ 0.002}_{- 0.002}$ &${\cal U} (0.01,20.00)$\\
13.     Emergence duration ${\cal I}_2$ (day)&$ 43.102 ^{+ 0.108}_{- 0.093}$ &$ 35.974 ^{+ 0.054}_{- 0.071}$ &${\cal U}_{\log} (0.000,200.000)$\\
14.     Decay duration ${\cal E}_2$ (day)&$ 49.676 ^{+ 0.042}_{- 0.058}$ &$18.087 ^{+ 0.019 }_{- 0.023}$ &${\cal U}_{\log} (0.000,200.000)$\\
(Third spot) &  &  & \\
15.     Latitude $\Phi_3$  (deg) &- &  $ 48.13 ^{+ 0.06}_{- 0.07}$
 &${\cal U} (\Phi_1,90.00)$\tablenotemark{{\rm c}}\\
16.     Initial longitude $\Lambda_3$ (deg)  &-&$ -72.37 ^{+ 0.26}_{- 0.25 }$
&${\cal U}  (-180.00,180.00)$\\
17.     Reference time $t_{3}$ (day) &- &$ 17.945 ^{+ 0.002}_{- 0.002}$
 &${\cal U} (0.00,26.229)$ \\
18.     Maximum radius $\alpha_{{ \textrm{max,3}}}$ (deg) &- &$ 10.514 ^{+ 0.004}_{- 0.004}$&${\cal U}(0.01,20.00)$\\
19.      Emergence duration ${\cal I}_3$ (day)&- &$ 3.915 ^{+ 0.004 }_{- 0.005}$ &${\cal U}_{\log} (0.00,200.000)$\\
20.     Decay duration ${\cal E}_3$ (day)&- &$ 24.680 ^{+ 0.036}_{- 0.064}$ &${\cal U}_{\log} (0.00,200.000)$\\
\hline
(Derived values) & &  & \\
Equatorial rotational velocity $\Omega_{\rm eq}$ (rad day$^{-1}$)&$2.3793^{+0.0001}_{-0.0001}$ &$ 2.3999 ^{+ 0.0003 }_{- 0.0001 }$ & \\
Rotational shear $\Delta \Omega$ (rad day$^{-1}$)&$0.0000^{+0.0001}_{-0.0000}$  &$ 0.1795^{+ 0.0006 }_{- 0.0006 }$ & \\ \hline
Unspotted level & $1.0147^{+0.0001}_{-0.0001}$ & $1.0146^{+0.0001}_{-0.0001}$ & \\ \hline
Reduced chi-square & $ 6450.345 ^{+ 0.115 }_{ -0.014 }$ & $ 3815.792 ^{+ 0.847}_{ -0.544}$ & \\
Logarithm of Model evidence $\log{\cal Z}$ &$-2260569.167$ & $-1322327.957$ & \\
\enddata
\tablenotetext{{\rm a}}{
${\cal U_{\text{log}}}(a,b)= 1/(\theta \log(b/a))$, ${\cal U}(a,b)=1/(b-a)$, ${\cal N}(\mu,\sigma^2)$, and ${\cal N}_{\rm log}(\mu,\sigma^2) = (1/\theta) \times {\cal TN}(\log \mu,(\sigma/\mu)^2)$  represent the bounded log uniform distribution (Jeffreys prior distribution) and bounded uniform distribution defined in $a\leq \theta \leq b$, and normal distribution and log-normal distribution with the mean $\mu$ and standard division $\sigma$ defined in $0 \leq \theta $, respectively.
}
\tablenotetext{{\rm b}}{The parameters in the spotted model correspond to $P_{\rm eq} = 2 \pi/ \Omega_{\rm eq}$ and $\kappa = \Delta \Omega/  \Omega_{\rm eq}$ (Equation \ref{eq:rot}), and we set normal prior distributions for $P_{\rm eq}$ and $\kappa$ with an error propagation in Table \ref{tb:stellar}.
}
\tablenotetext{{\rm c}}{
We discern each spot by its latitude $\Phi_k$ as in \cite{Ikuta23}. In the case of the two-spot model, we set $\Phi_3 = 90.0$ (the upper limit of the latitude).
}
\end{deluxetable*}

\subsection{Calculation of the B-band light curve} \label{sec:bband}
We reproduce the B-band light curve by converting limb-darkening and spot intensity (Table \ref{tb:stellar}) using deduced stellar and spot parameters obtained from the starspot mapping for the TESS light curve (Table \ref{tb:para}).
We compare the reproduced light curves from the two- and three-spot models with B-band light curve in the Sector 50 (after March 26 in 2022) since the B-band light curve was sparsely obtained (Figure \ref{fig:lightcurve}).
We note that an intensive photometry in bluer bands than that in the TESS one enables to break some degeneracies between spot intensity and size and to constrain the parameters of the starspot mapping only from the TESS light curve  \citep[][]{Mori24}.

\subsection{Calculation of the light curve from Zeeman Doppler Imaging} \label{sec:dilc}
We reproduce the light curve from the intensity map of the ZDI \citepalias{Namekata24} to compare it with the TESS light curve.
The ZDI reconstructs the intensity map on the discretized stellar surface as the intensity contribution of each surface element, and is different from the starspot mapping with circular spots \citep[][]{Ikuta20}.
First, we derive the temperature of each surface element, where $R(\lambda)=1$ and the wavelength range of 470 to 1050 nm (Section \ref{sec:obs}), and we calculate the relative intensity $f_{\rm int}$ in the TESS-band for each, based on Equation \ref{eq:relspot}.
We note that we treat a region with intensity larger than one as a bright spot with intensity independent of the viewing angle, unlike stellar faculae.
Second, we calculate the relative flux $F(t)/F_{\rm ave}$ from the modeled flux \citep{kipping12} divided by its average in time:

\begin{align} \label{eq:flux}
F (t) &= 1- \sum_{n=0}^4 \Biggl(\frac{nc_n}{n+4}\Biggr) \notag \\ &- \frac{1}{\pi} \int \int_{\cos \beta (t) \geq 0} d\Phi d\Lambda \cos \Phi \cos \beta (t)  \notag \\ &\times (1-f_{\rm int}) \biggl[1-\sum_{n=1}^4 c_n \{1-\cos^{\frac{n}{2}}\beta (t) \} \biggr],
\end{align}
where 
\begin{align} \label{eq:colon}
\cos \beta (t) &= \cos i \sin \Phi \notag \\ &+ \cos \Phi \sin i \cos ( \Lambda+ 2\pi t/P_{\rm rot} )
\end{align} 
is cosine of the angle between the line of sight and the line from the center of the star to the spot center \citep[for details,][]{Ikuta20} at the time $t$ from the reference time of the ZDI (Section \ref{sec:obs}).In addition, the dark and bright area weighted with the intensity are calculated with 
\begin{equation} \label{eq:area}
S (t) = \frac{1}{\pi} \int \int_{\cos \beta (t) \geq 0} d\Phi d\Lambda \cos \Phi \cos \beta (t) f_{\rm int},
\end{equation}
under the condition of $f_{\rm int}<1$ and $f_{\rm int}\geq1$, respectively.

\begin{figure}[tbhp!]
\plotone{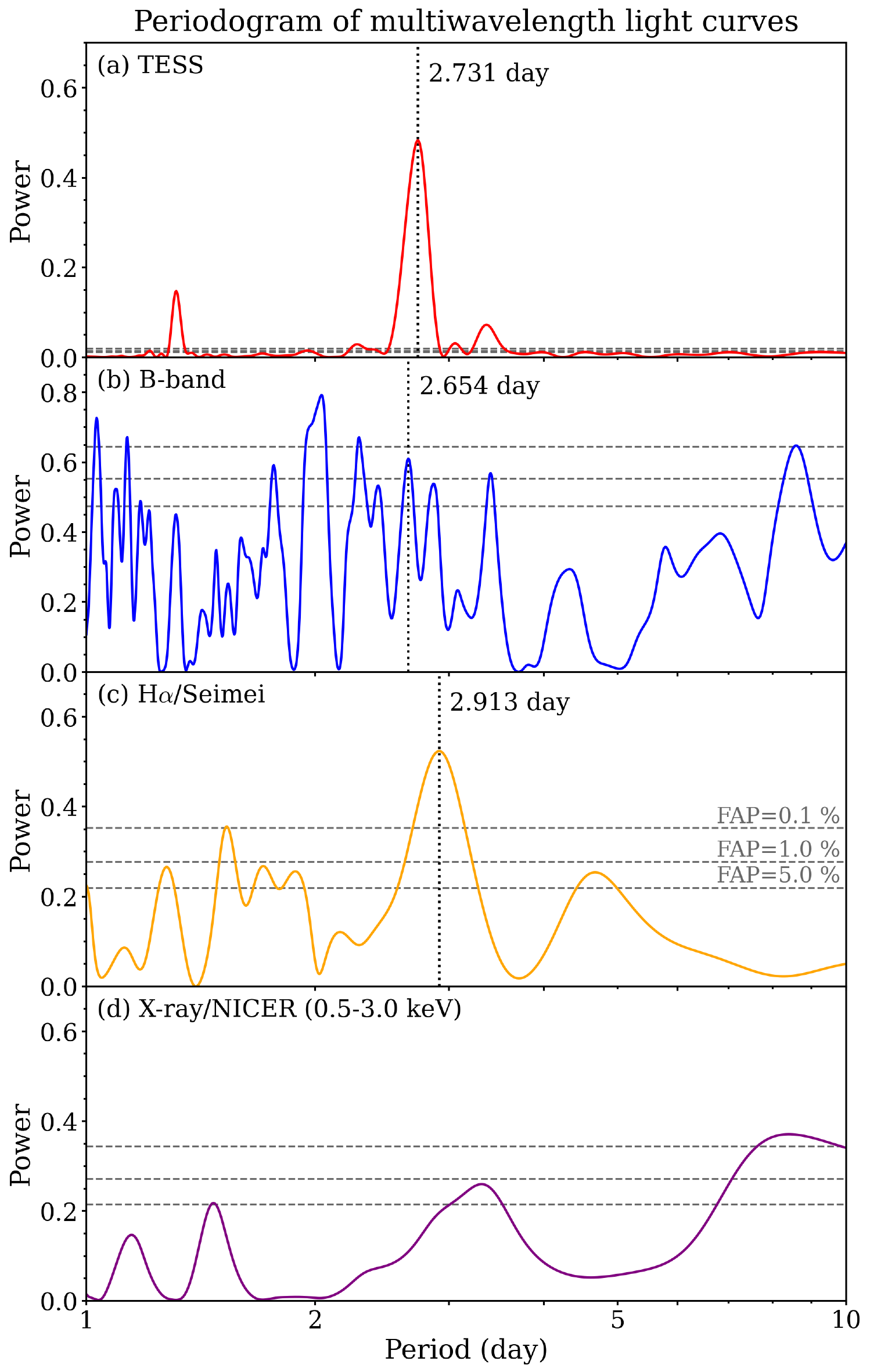} 
\caption{The GLS periodogram for multiwavelength light curves of the (a) TESS, (b) B-band, (c) H$\alpha$, and (d) X-ray (red, blue, orange, and purple). 
The vertical lines show the periodicity of the TESS, B-band, H$\alpha$ light curves for
2.731, 2.654, and 2.913 days under the false alarm probability (FAP) smaller than 1\%.
}
 \label{fig:period}
\end{figure}

\begin{figure}[tbhp!]
\plotone{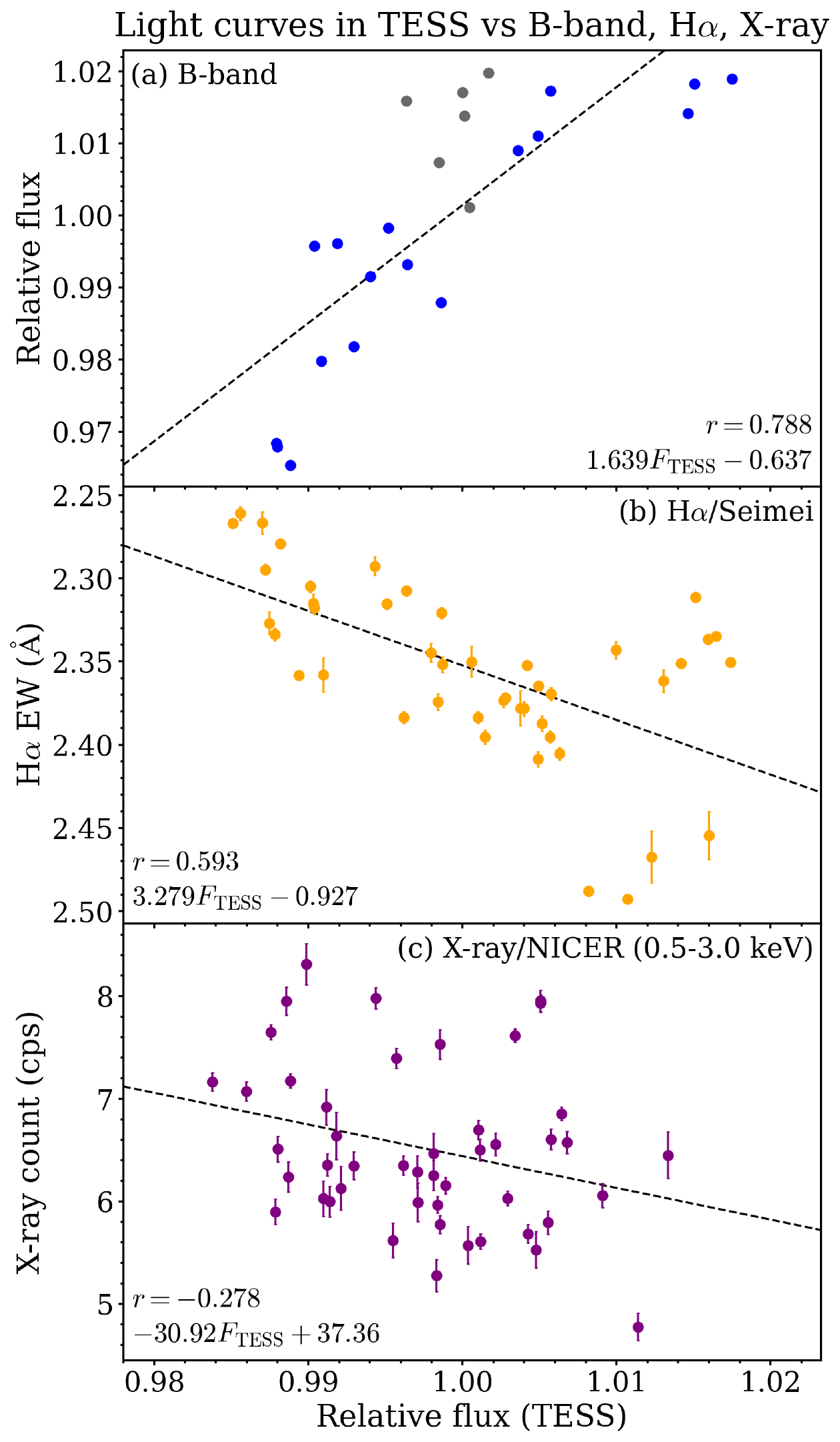} 
\caption{The light curves of (a) B-band, (b) H$\alpha$, and (c) X-ray versus the TESS light curve (blue, orange, and purple). The black lines derived from the least squares are formulated by $1.639F_{\rm TESS}-0.637$,  $3.279F_{\rm TESS}-0.927$, and $-30.92F_{\rm TESS}+37.36$, where $F_{\rm TESS}$ is the relative flux of the TESS, respectively. The Pearson correlation coefficients are calculated to be $r=0.788$, $0.593$, and $-0.278$, respectively.
We note that the H$\alpha$ emission is defined to be negative, and the vertical axis is reversed to clarify the emission in the panel (b).
}
 \label{fig:flux}
\end{figure}

\subsection{Period analysis} \label{sec:period}
We perform the period analysis of the TESS, B-band, H$\alpha$, and X-ray light curves (Figure \ref{fig:lightcurve}), with the Generalized Lomb-Scargle periodogram \citep{GLS}, after flares and outliers are masked.
As a result, we obtain the periodicity of $
2.731 \pm  0.006$,  $2.654 \pm 0.021$, and $2.913\pm0.071$ days with the false alarm probability smaller than $1$ \%, for the TESS, B-band, and H$\alpha$ light curves, respectively (Figure \ref{fig:period}). 
We cannot detect the periodicity in the X-ray light curve although the rotational phase is almost covered.
In addition, we calculate the Pearson correlation coefficients between the TESS light curve and B-band, H$\alpha$, and X-ray light curves, to investigate the linear correlation \citep[e.g.,][]{Wargelin24}. 
As a result, each of the coefficient is calculated to be 0.788, 0.593, and -0.278, respectively (Figure \ref{fig:flux}).  
Therefore, the B-band light curve is mostly correlated with the TESS light curve. The H$\alpha$ light curve is almost correlated with the TESS light curve such that the H$\alpha$ emission in the H$\alpha$ EW becomes larger when the relative flux in the TESS light curve becomes smaller. We note that the H$\alpha$ emission is defined to be negative (Section \ref{sec:kools}).
The X-ray light curve is weakly anti-correlated with the TESS light curve.
These results support the above period analysis.

\subsection{Multidimensional Gaussian process for multiwavelength data} \label{sec:multigp}
We jointly regress the TESS, B-band, H$\alpha$, and X-ray light curves in the second half of Sector 50 with a Gaussian process \citep[GP; for a review,][]{Aigrain23} under an assumption of all data with the same degree of the time scale in the variability to complement the sparse time-series data.
This is because the contributions of active regions to the H$\alpha$ and X-ray light curves are fully unexplored, and the result of the starspot mapping can only reproduce the overall features of the TESS light curve.
The multidimensional GP has been implemented for contemporaneous multicolor photometry 
\citep{Gordon20}, multiple data of radial velocity and activity indicators \citep{Rajpaul15,Delisle22,Barragan22}, and those with photometry \citep{Tran23}.
We also introduce the multidimensional GP for multiwavelength data on different time and its cadence by utilizing \texttt{tinygp} \citep{tinygp}. Then, we adopt the modified quasi-periodic kernel for efficient computation linear with the number of data \citep{celerite}:
\begin{align}
k(|t_{a,i}-t_{b,j}|) &= \sigma^2_{{\rm jit},a} \delta_{ab} \delta_{ij}  \notag \\
&+ \frac{A_a A_b}{2+B} \exp \{ -C|t_{a,i}-t_{b,j}| \} \notag \\ &\times \{ \cos \{ \frac{2\pi |t_{a,i}-t_{b,j}|}{P} \} + (1+B) \},
\end{align}
where $(a,b)$ denotes the label of the band or wavelength of the TESS, B-band, H$\alpha$, and X-ray, and $(i,j)$ denotes the number of data point for each.
$A_a$, $B$, $C$, and $P$ are hyperparameters for the amplitude, constant scale, length scale, and periodicity, of the kernel, respectively.  
The X-ray light curve is expected to be anti-correlated with the TESS and B-band light curves, and we take negative values for the parameter $A_{\rm X}$. 
We note that the H$\alpha$ emission is expected to be correlated with the TESS and B-band light curves because we define the emission to be negative (Section \ref{sec:period}).
We also adopt parameters $\gamma_a$ as the mean of Gaussian process for the offsets of the variability \citep[e.g.,][]{Wargelin24}.
We perform the regression of the multiwavelength light curve with the No U-turn sampler \citep{NUTS} as an efficient sampling by a Hamiltonian Monte Carlo implemented in \texttt{NumPyro} \citep{numpyro}.

\section{Result and Discussion} \label{sec:result}

\begin{figure*}[tbhp!]
\plotone{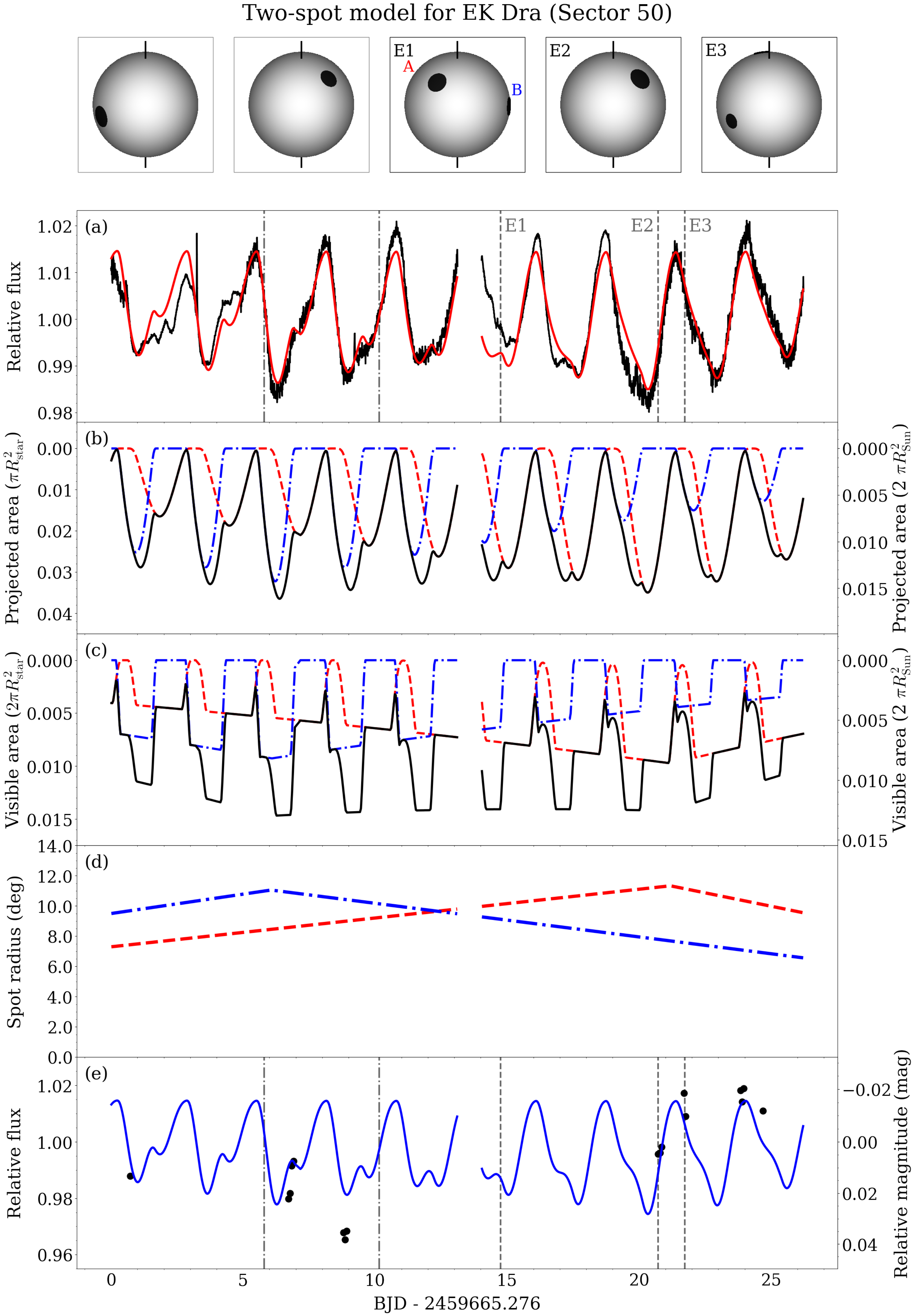}
\caption{
(Top) Spot maps at the different rotational phases including the E1, E2, and E3, for the two-spot model. (Bottom) (a) the TESS light curve (black) and reproduced one (red). (b) the temporal variation of visible projected area of each spot (red and blue) and the total (black) relative to the stellar disk and solar hemisphere, (c) that of visible area relative to the stellar and solar photospheres, (d) the temporal variation of each spot radius (red and blue), and (e) the B-band light curve (black) and light curve reproduced with parameters derived from the two-spot model (blue).
} \label{fig:S2_spotmap}
\end{figure*}

\begin{figure*}[tbhp!]
\plotone{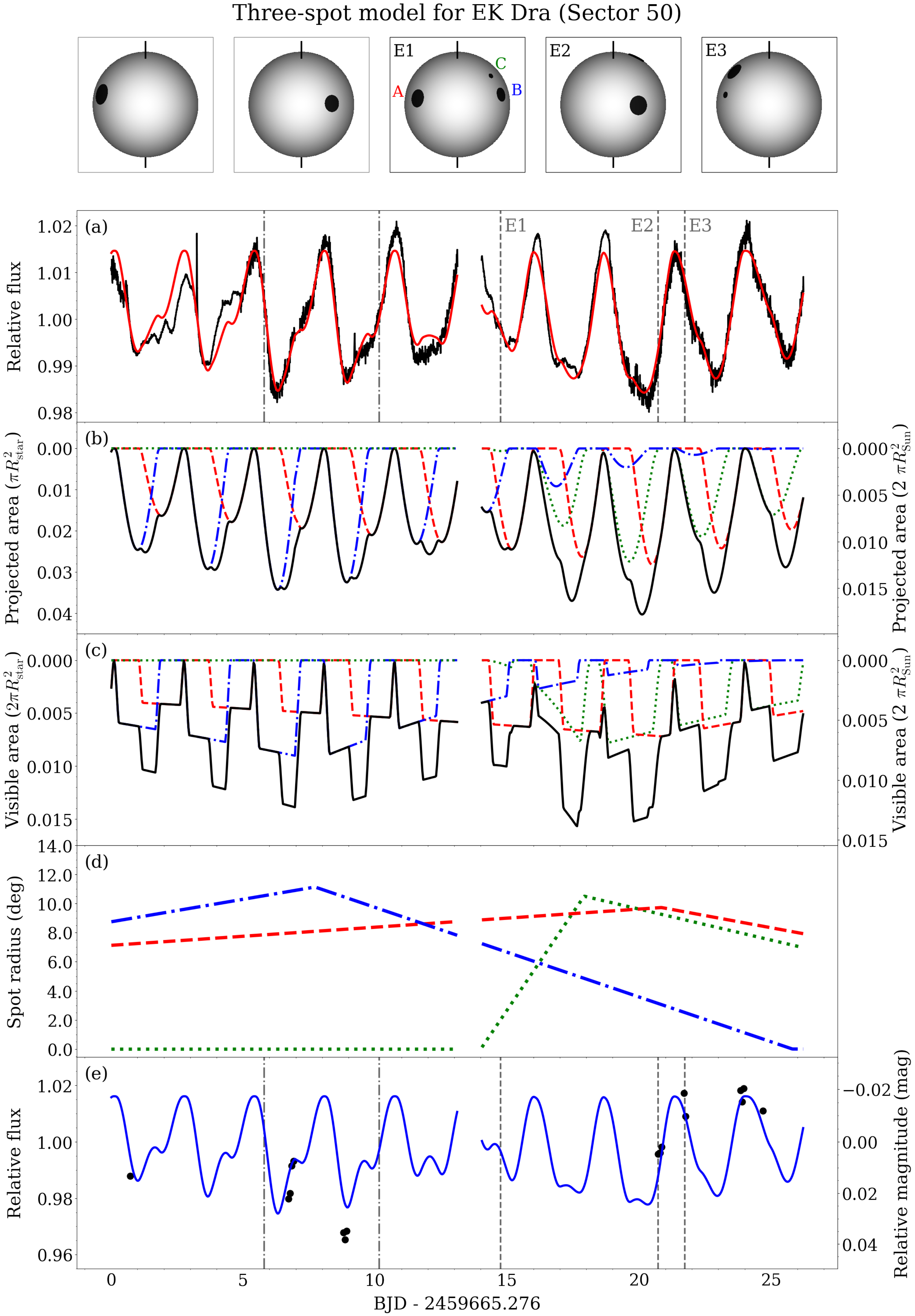}
\caption{Same as Figure \ref{fig:S2_spotmap}, but for the three-spot model (red, blue, and green).
} \label{fig:S3_spotmap}
\end{figure*}

We optimize the TESS light curves of EK Dra in the Sector 50 with the two- and three-spot models (Section \ref{sec:spotmap}).
Table \ref{tb:para} exhibits deduced parameters with their credible regions and the logarithm of the model evidence for each of the model, together with their prior distributions for each of the parameters. 
Figure \ref{fig:S2_spotmap} and \ref{fig:S3_spotmap} exhibit visualized surface at different phases for two- and three-spot models, respectively. Each of the figure also show (a) the TESS and reproduced light curves (black and red), (b) the temporal variation of visible projected area of each spot (red, blue, and green) and the total (black), (c) that of visible area, (d) that of each spot radius (red, blue, and green), and (e) B-band light curve (black) and light curve reproduced with deduced parameters (blue) (Section \ref{sec:bband}). The three-spot model is preferred from the model comparison because the likelihood becomes larger by adopting more parameters and compensating for the residuals of the two-spot model. 
We note that the unspotted level is derived as the maximum of the relative flux for both models (i.e., any of the spots are out of view at the maximum), because we adopt only two or three spots to reproduce the overall features of the light curve. The unspotted level would be larger than that from two- and three-spot models if more spots were adopted, although it may be difficult to determine the unspotted level only from the monochromatic light curve (Section \ref{sec:spotmap}).

\begin{deluxetable}{lc}[tbhp!]
\tablecaption{Deduced hyperparameters of GP for multiwavelength variability \label{tb:multigp}}
\tabletypesize{\scriptsize}
\tablehead{
\colhead{Parameters} &  \colhead{Derived values} 
}
\startdata
(Mean of GP) &  \\ 
1. Offset $\gamma_{\rm TESS}$ & $ 1.003 ^{+ 0.010 }_{- 0.010 }$  \\ 
2. Offset $\gamma_{\rm B}$ & $ 1.007^{+ 0.009 }_{- 0.009 }$ \\ 
3. Offset $\gamma_{\rm H \alpha}$ &$ 2.362^{+ 0.031 }_{- 0.030}$\\ 
4. Offset $\gamma_{\rm X}$ &$ 6.478 ^{+ 0.157 }_{- 0.154 }$  \\ 
(Jitter term) &  \\
5. Jitter $\log \sigma_{{\rm jit, TESS}}$ &  $ -14.454 ^{+ 3.770 }_{- 3.817 }$ \\
6. Jitter $\log \sigma_{{\rm jit, B}}$ &$ -5.692 ^{+ 0.279 }_{- 0.292 }$ \\
7. Jitter $\log \sigma_{{\rm jit, H\alpha}}$ &  $ -3.135^{+ 0.108 }_{- 0.108}$\\
8. Jitter $\log \sigma_{{\rm jit, X}}$ & $ -0.238 ^{+ 0.106 }_{- 0.106 }$  \\
(Modified quasi-periodic term) &  \\
9. Amplitude $\log A_{{\rm TESS}}$ & $ -4.099 ^{+ 0.463}_{- 0.447 }$   \\
10. Amplitude $\log A_{{\rm B}}$ & $ -4.291^{+ 0.534 }_{- 0.425 }$ \\
11. Amplitude $\log A_{{\rm H\alpha}}$ &$ -3.013 ^{+ 0.525 }_{- 0.505 }$  \\
12. Amplitude $\log (-A_{{\rm X}})$ &$ -7.544 ^{+ 6.574 }_{- 8.006 }$  \\
13. Constant $\log B$ &   $ -11.033 ^{+ 6.088 }_{- 6.228 }$  \\
14. Exponential length $\log C$ &  $ -3.178^{+ 0.900 }_{- 0.928 }$ \\
15. Periodicity $\log P$ & $ 0.949 ^{+ 0.034 }_{- 0.034 }$ \\
\enddata
\end{deluxetable}

\begin{figure*}[tbhp!]
\plotone{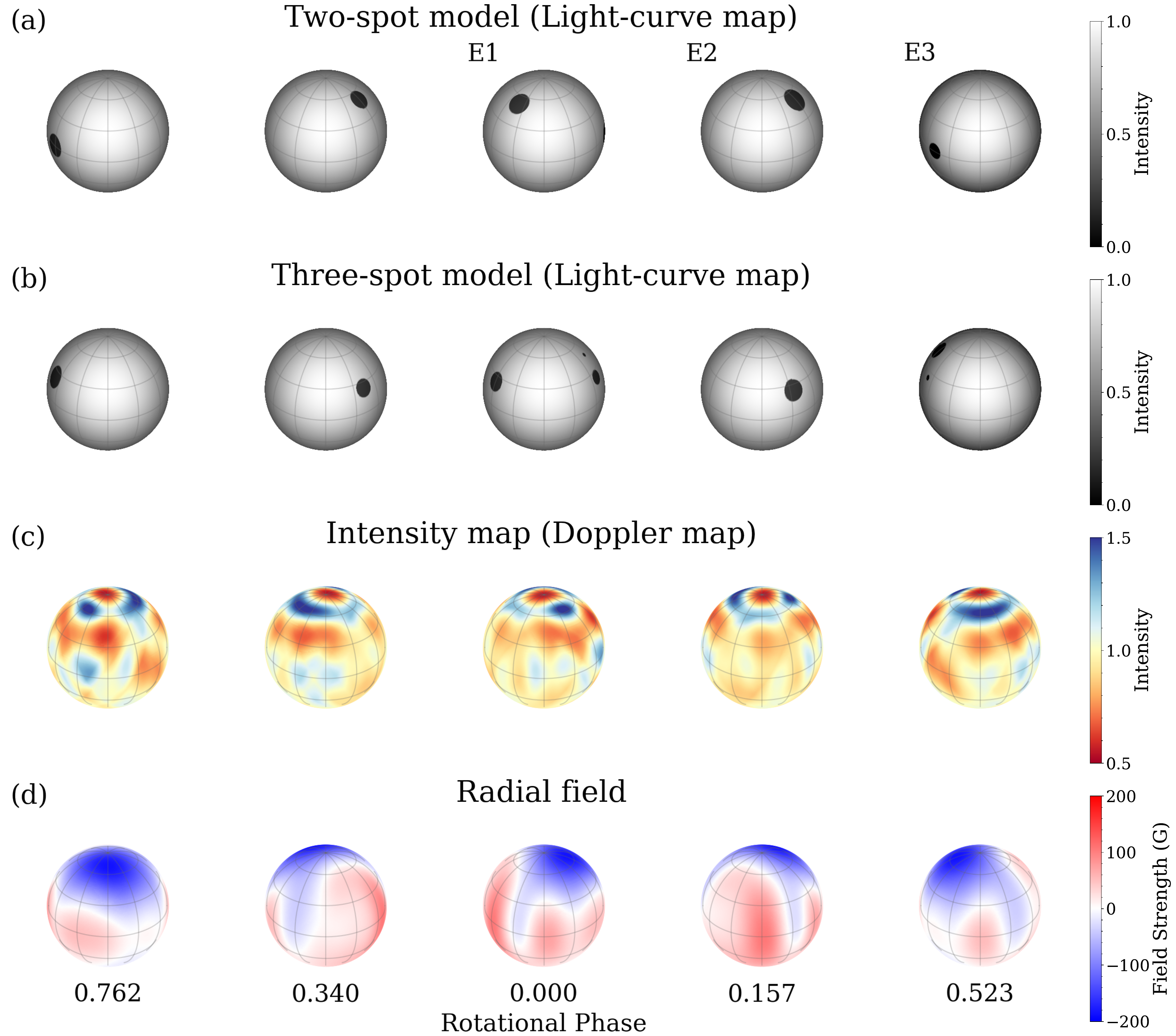}
\plotone{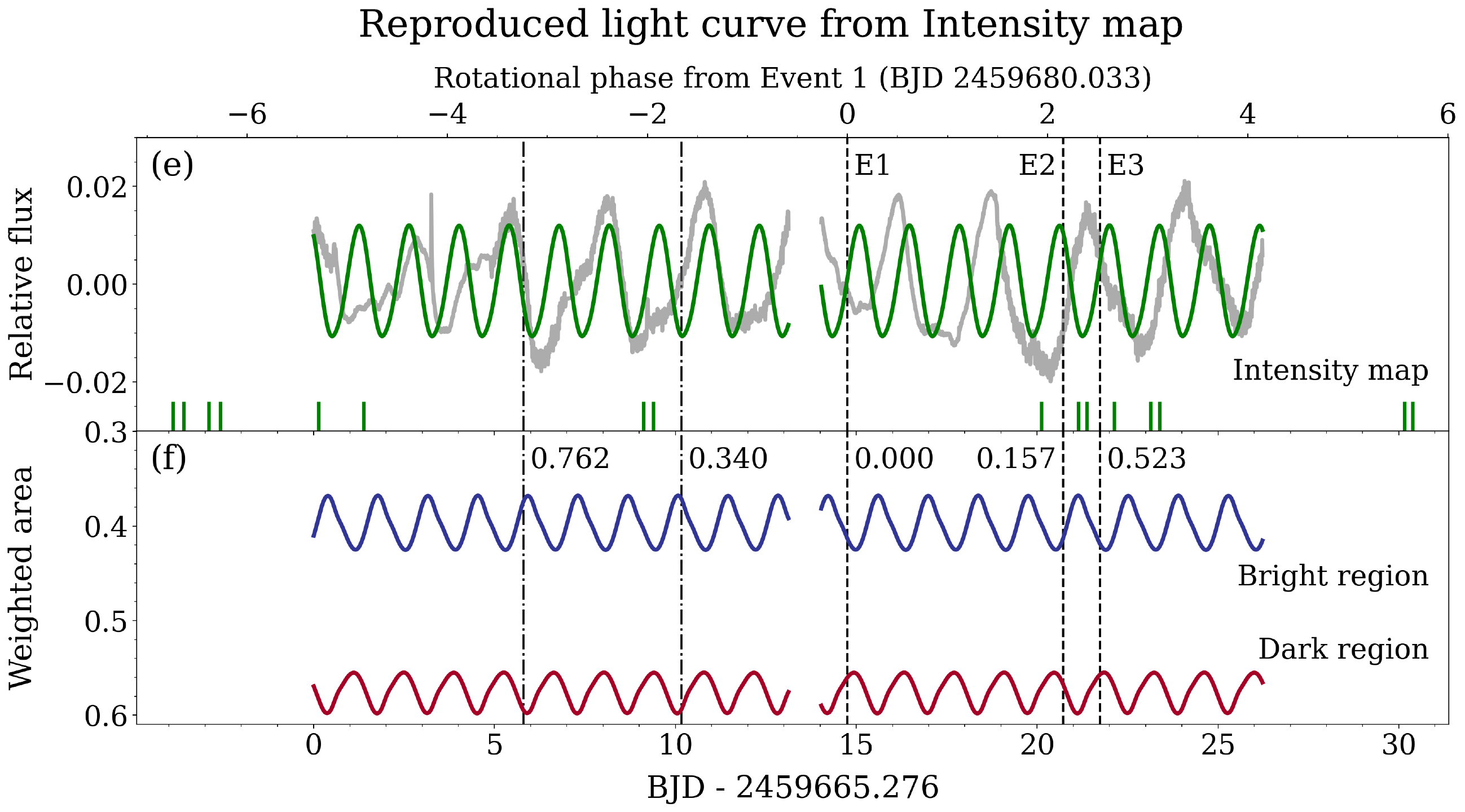}
\caption{
(a,b) Spot maps from the two- and three spot models (Light-curve map), (c) intensity map (Doppler map), and (d) magnetic map of the radial field, both from the ZDI (Section \ref{sec:obs}), at different rotational phases including the E1, E2, and E3. (e) The TESS light curve (gray) and light curve reproduced with the intensity map (green) and (f) the rotational variations of the area weighted with the intensity (weighted area) for bright and dark regions in the intensity map (blue and red) (Section \ref{sec:dilc}).
The rotational phases from the E1 are 0.762, 0.340, 0.000, 0.157, and 0.523 (black), respectively.} \label{fig:compare_doppler}
\end{figure*}

\begin{figure*}[tbhp!]
\plotone{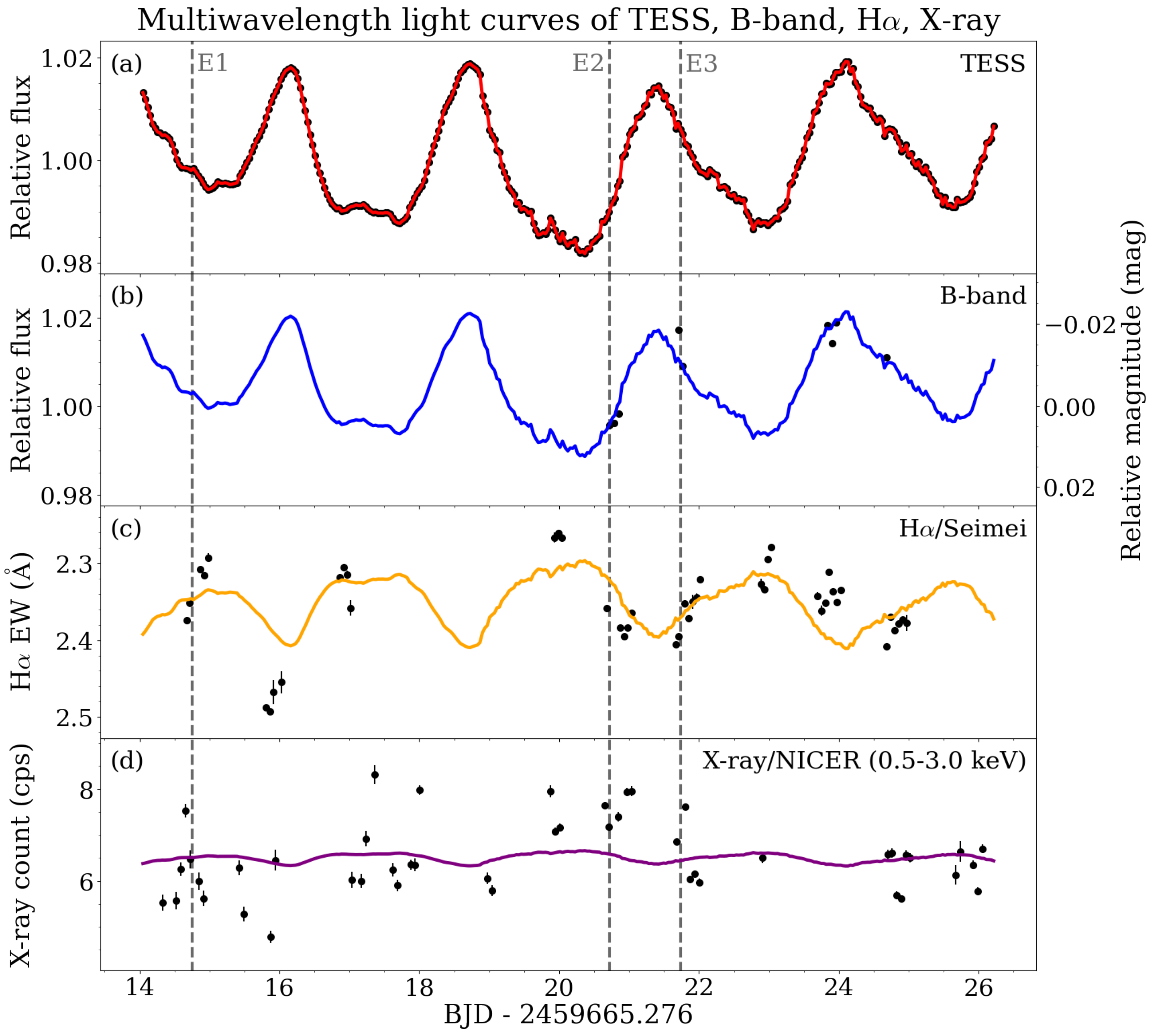}
\caption{Light curves of the TESS, B-band, H$\alpha$, and X-ray, during the second half of the Sector 50 (black). The light curves are jointly optimized with multidimensional GP to compare the multiwavelength variability ascribed to spots and active regions (Section \ref{sec:multiwave}). The light curves are reproduced from the posterior distribution marginalized with hyperparameters (red, blue, orange, and purple). We note that the H$\alpha$ emission is defined to be negative, and the vertical axis is reversed to clarify the emission in the panel (c).
} \label{fig:TESS_X}
\end{figure*}

Figure \ref{fig:compare_doppler} exhibits the comparison of surface map at different rotational phases between (a,b) the models from the TESS light curve (hereafter, Light-curve map), (c) intensity map (hereafter, Doppler map), and (d) radial magnetic field, both from the ZDI, (e) the reproduced light curve from the Doppler map (green) (Section \ref{sec:dilc}) and (f) the area weighted with the intensity for bright and dark regions in the Doppler map (blue and red). 
Table \ref{tb:multigp} lists deduced parameters of the GP, and Figure \ref{fig:TESS_X} shows multiwavelength variability in the TESS, B-band, H$\alpha$, and X-ray light curves, and reproduced light curves from the posterior distribution marginalized with hyperparameters in the multidimensional GP (Section \ref{sec:multigp}).
We discuss the result in comparison with previous studies by the ZDI (Section \ref{sec:di}), multiwavelength variability in the TESS, B-band, H$\alpha$, and X-ray (Section \ref{sec:multiwave}), the relation between spot locations and superflares (Section \ref{sec:timing}).

\subsection{Comparison between spot mapping and Zeeman Doppler Imaging} \label{sec:di}
We compare the Light-curve map with the Doppler map. In Figure \ref{fig:S2_spotmap}, \ref{fig:S3_spotmap}, and \ref{fig:compare_doppler}, either of the spot in both of the Light-curve maps and Doppler map is visible to the line of sight in all the rotational phases.
Spots are deduced to be on the equator and mid-latitude from the spot mapping for both of the models, and these spots in the Light-curve maps partially correspond to the dark regions in the Doppler map, although the Doppler map exhibits extended dark regions. There are also no polar spot in the Light-curve maps, but there is a polar dark region in the Doppler map. 
This is probably because a polar spot gives small contribution to the modulation of the light curve, and we assume only two and three spots to reproduce the overall structure of the TESS light curve, while a polar spot can be reconstructed from the shape of the spectral lines in the ZDI.

In Figure \ref{fig:compare_doppler}, we also compare the reproduced light curve from the Doppler map with the TESS light curve. 
The amplitude from the Doppler map results in being less than the TESS light curve because the amplitude is reflected with the asymmetry of the spot distribution, and the Doppler map is obtained by being averaged in the observation period ($\sim 34$ day). 
In addition, the coverage of the rotational phase in the DI does not dense enough to determine the differential rotation at the time of observation, while the significant differential rotation is reported on EK Dra \citep{Waite17}. The differential rotation would also progressively distorts the spot distribution.
We note that the result of spot locations from the photometry and DI would be different due to the methods of the mapping and imaging 
 \citep[][]{Luger21c}.
Therefore, joint dynamical mapping from photometry and spectroscopy (Doppler Imaging) with variable spot size and differential rotation is required to obtain more robust spot map in future works (Section \ref{sec:conclusion}). We also note that the spatial resolution of the magnetic field is low ($\simeq$ 25 deg), and only large-scale magnetic fields are reconstructed because small-scale fields are canceled out.
We compare the result of the spot mapping with the previous results of the DI for EK Dra in the past three decades. Table \ref{tb:doppler} summarizes spot properties obtained from the DI and ZDI in the previous studies.  The result of the spot mapping for the models shows the spots near the equator to mid-latitude, and this result is consistent with the previous studies \citep[][\citetalias{Namekata24}]{Strassmeier98, Jarvinen07,Rosen16, Waite17, Jarvinen18, Senavci21} for the overall spot latitudes.

\subsection{Multiwavelength variability} \label{sec:multiwave}

We compare the multiwavelength variability ascribed to spots and active regions observed in the TESS, B-band, H$\alpha$, and X-ray light curves.
In Figure \ref{fig:S2_spotmap} and \ref{fig:S3_spotmap}, the light curves reproduced from the deduced parameters of the spot mapping in the B-band almost correspond to the B-band light curve except for the data from 8.80 to 8.92 (BJD - 2459665.276).
In Figure \ref{fig:TESS_X}, the regression with the GP is mostly determined from the TESS light curve because of its small uncertainties and of sparse light curves except for the TESS light curve. The regressed light curve almost corresponds to the B-band light curve, and is almost synchronized with the H$\alpha$ light curve. However, some data in the H$\alpha$ significantly deviate from the regressed light curve probably because the H$\alpha$ emission is radiated from the chromosphere above active regions with plage. 
In the solar case, the activity indicators exhibit rotational variations ascribed to active regions \citep{Maldonado19}, and the rotational variation of Balmer lines from the solar chromosphere on active regions almost corresponds to the Total Solar Irradiance \citep{Criscuoli23}. The result is also consistent with the solar case and supports rotational modulations of the H$\alpha$ light curve for M-dwarfs \citep{Namekata20,Maehara21,Schoefer22,Odert25,Notsu25} and a solar-type star EK Dra \citep{Namekata22b}.
Therefore, the H$\alpha$ light curve can be indicators of active regions corresponding to the photosphere.
The intensity of the H$\alpha$ spectrum would be reflected for the type of sunspots \citep{Toriumi19}, and we need to explore the relation between the magnetic flux and H$\alpha$ spectrum simultaneously in future works. 

For the X-ray light curve, there are large scatter as well as the fact that clear periodicity cannot be detected from the periodogram (Section \ref{sec:period}), although \cite{Guedel95} has reported a clear periodicity of EK Dra at its activity minimum (November 1990) \citep{Dorren94}. This is also supported by the long-term photometric observations \citep{Jarvinen18}, suggesting that the observation period in this study (April 2022) was close to the expected activity maximum from the activity cycle of EK Dra ($\sim$ 9 years). Moreover, sunspots in active regions also exhibit the rotational variations in X-ray associated with the heating of solar coronae through the dissipation of magnetic fields \citep[for a review,][]{Gudel04}, and the power of the periodicity depends on the activity level \citep[e.g.,][]{Ueno97,Takeda19}. 
Spatial structure of the corona above sunspots also contribute to the X-ray light curve \citep{Toriumi20}.
It is also reported that there is a periodic modulation in the X-ray light curve for a young K-dwarf AB Doradus \citep{Hussain05}. The X-ray is suggested to be radiated from compact coronae on active regions, and short-term scatters in the X-ray light curve may be due to flares.
\cite{Hussain07} suggests that the X-ray light curve depends on
the distribution of magnetic fields from the contemporaneous ZDI.
Moreover, a long-term variability in optical, UV, and X-ray has been investigated for an active mid M-dwarf Proxima Centauri \citep[Pro Cen;][]{Wargelin17,Wargelin24}.
The optical light curve for Pro Cen is strongly and weakly anticorrelated with the UV and X-ray light curves, respectively.
\cite{Wargelin24} explains that this is because the X-ray is radiated from the
optically thin and spatially extended corona whereas the optical and UV emission are from the photosphere and chromosphere.
Therefore, we conclude that the scattered X-ray light curve of EK Dra may reflect multiple spots on high activity level and be contaminated by small flares.

\begin{deluxetable*}{lcccccc}[tbhp!]
\tablecaption{Spot parameters from Sector 50 at the time of Event 1, 2, and 3$^{*}$ \label{tb:prom}}
\tabletypesize{\scriptsize}
\tablehead{
\colhead{} &  \multicolumn{2}{c}{Event 1}  &  \multicolumn{2}{c}{Event 2}  & \multicolumn{2}{c}{Event 3} \\
\colhead{Parameters} &  \colhead{Two-spot Model} & \colhead{Three-spot Model$^{\rm a}$}  &  \colhead{Two-spot Model} & \colhead{Three-spot Model} &  \colhead{Two-spot Model} & \colhead{Three-spot Model}
}
\startdata
(Spot A)& &  & &  & & \\
Longitude   (deg)$^{\rm b}$&  $ -37.87 ^{+ 0.03 }_{- 0.03}$ & $ -56.84 ^{+ 0.01 }_{- 0.06}$ &$ 55.31^{+ 0.03}_{- 0.03}$ & $ 32.37 ^{+ 0.06}_{- 0.06 }$  & $ -166.46 ^{+ 0.02 }_{- 0.04}$ & $ 169.94 ^{+ 0.01 }_{- 0.06}$\\
Colongitude (deg)$^{\rm c}$&$ -25.28 ^{+ 0.03 }_{- 0.03 }$ &$ -49.81 ^{+ 0.02 }_{- 0.02 }$ &${\bf  36.57 ^{+ 0.02 }_{- 0.02}}$ &${\bf 29.01^{+ 0.03 }_{- 0.03}}$  &$ -134.05^{+ 0.19 }_{- 0.21 }$ & $ 164.20 ^{+ 0.06 }_{- 0.06 }$ \\ 
Radius  (deg) &$ 10.102 ^{+ 0.004 }_{- 0.005}$ &  $ 8.969 ^{+ 0.002 }_{- 0.002 }$  &$ 11.238 ^{+ 0.004}_{- 0.004}$&$ 9.713 ^{+ 0.004 }_{- 0.003 }$ & $ 11.147 ^{+ 0.005}_{- 0.005}$ & $ 9.444 ^{+ 0.005 }_{- 0.004 }$\\
Radius   ($10^{10}$ cm) &$ 1.17 ^{+ 0.07}_{- 0.10}$ &  $ 1.02 ^{+ 0.08}_{- 0.07}$ & $ 1.29 ^{+ 0.09 }_{- 0.10 }$ & $ 1.11^{+ 0.08 }_{- 0.09}$ &$ 1.28 ^{+ 0.09 }_{- 0.10}$  &  $ 1.08 ^{+ 0.07 }_{- 0.09 }$\\
(Spot B)& &  & &  & & \\
Longitude (deg)$^{\rm b}$&$ 82.64 ^{+ 0.02}_{- 0.04}$ &$ 67.15 ^{+ 0.06 }_{- 0.01}$& $ 175.81 ^{+ 0.05}_{- 0.03 }$&$ 156.42 ^{+ 0.01}_{- 0.06}$  &  $ -45.96 ^{+ 0.03 }_{- 0.05 }$ & $ -66.06 ^{+ 0.06 }_{- 0.06 }$ \\ 
Colongitude (deg)$^{\rm c}$&  ${\bf 82.61 ^{+ 0.05 }_{- 0.05 }}$ & ${\bf 58.25 ^{+ 0.02 }_{- 0.03 }}$ & $ 175.07 ^{+ 0.05 }_{- 0.05 }$& $ 144.41 ^{+ 0.05 }_{- 0.05}$ & ${\bf -49.25 ^{+ 0.05 }_{- 0.05} }$ &$ -57.36 ^{+ 0.06 }_{- 0.05 }$  \\ 
Radius  (deg) & $ 9.124 ^{+ 0.004 }_{- 0.004}$  &$ 6.828 ^{+ 0.003}_{- 0.004}$  & $ 7.796 ^{+ 0.004 }_{- 0.004 }$& $ 3.153 ^{+ 0.007 }_{- 0.009}$ & $ 7.571 ^{+ 0.004}_{- 0.004}$  & $ 2.528 ^{+ 0.008 }_{- 0.009 }$  \\
Radius  ($10^{10}$ cm) & $ 1.04 ^{+ 0.08 }_{- 0.07 }$ &$ 0.79 ^{+ 0.05 }_{- 0.07}$ &$ 0.89 ^{+ 0.07}_{- 0.07}$ &  $ 0.36 ^{+ 0.02 }_{- 0.03}$   & $ 0.88^{+ 0.05}_{- 0.08}$& $ 0.29 ^{+ 0.02 }_{- 0.02 }$ \\
(Spot C)& &  & &  & & \\
Longitude  (deg)$^{\rm b}$ &-&$ 72.71 ^{+ 0.06}_{- 0.06}$ &- &$ 138.94 ^{+ 0.06 }_{- 0.06 }$  &- &$ -87.43^{+ 0.06 }_{- 0.06}$\\
Colongitude (deg)$^{\rm c}$&- & $ 49.48 ^{+ 0.09 }_{- 0.05}$ &- &$ 98.23 ^{+ 0.18 }_{- 0.13}$  &- &  ${\bf -59.14 ^{+ 0.03 }_{- 0.05}}$  \\ 
Radius  (deg) &-& $ 1.948 ^{+ 0.007 }_{- 0.006 }$ &- &$ 9.331 ^{+ 0.004 }_{- 0.004 }$  &- &  $ 8.899 ^{+ 0.003}_{- 0.004 }$\\
Radius ($10^{10}$ cm) &- &$ 0.22^{+ 0.02}_{- 0.02 }$ &- &  $ 1.07 ^{+ 0.07 }_{- 0.09}$  &- &  $ 1.01^{+ 0.08 }_{- 0.07}$\\
\enddata
\tablenotetext{$*$}{The colongitude of a spot accompanied by the superflare is marked (Section \ref{sec:timing}).}
\tablenotetext{{\rm a}}{The three-spot model at the time of Event 1 is adopted in \citetalias{Namekata24}.}
\tablenotetext{{\rm b}}{
The longitude for each of the spot latitude $\Phi_k$ at the time $t$ is calculated by $\Lambda_k + 2 \pi t/P(\Phi_k)$ (deg), where $P(\Phi_k)$ is given by  Equation \ref{eq:rot}. The radius at the time $t$ is calculated by Equation \ref{eq:spotrad} and transformed by multiplying the stellar radius in Table \ref{tb:stellar}.}
\tablenotetext{{\rm c}}{The tangent of the colongitude is given by $\sin \Lambda_t \cos \Phi/(\cos i \sin \Phi + \cos \Phi \sin i \cos \Lambda_t)$, where $\Lambda_t = \Lambda + 2 \pi t/P(\Phi)$ at the time of $t$.}
\end{deluxetable*}

\begin{figure*}[tbhp!]
\plotone{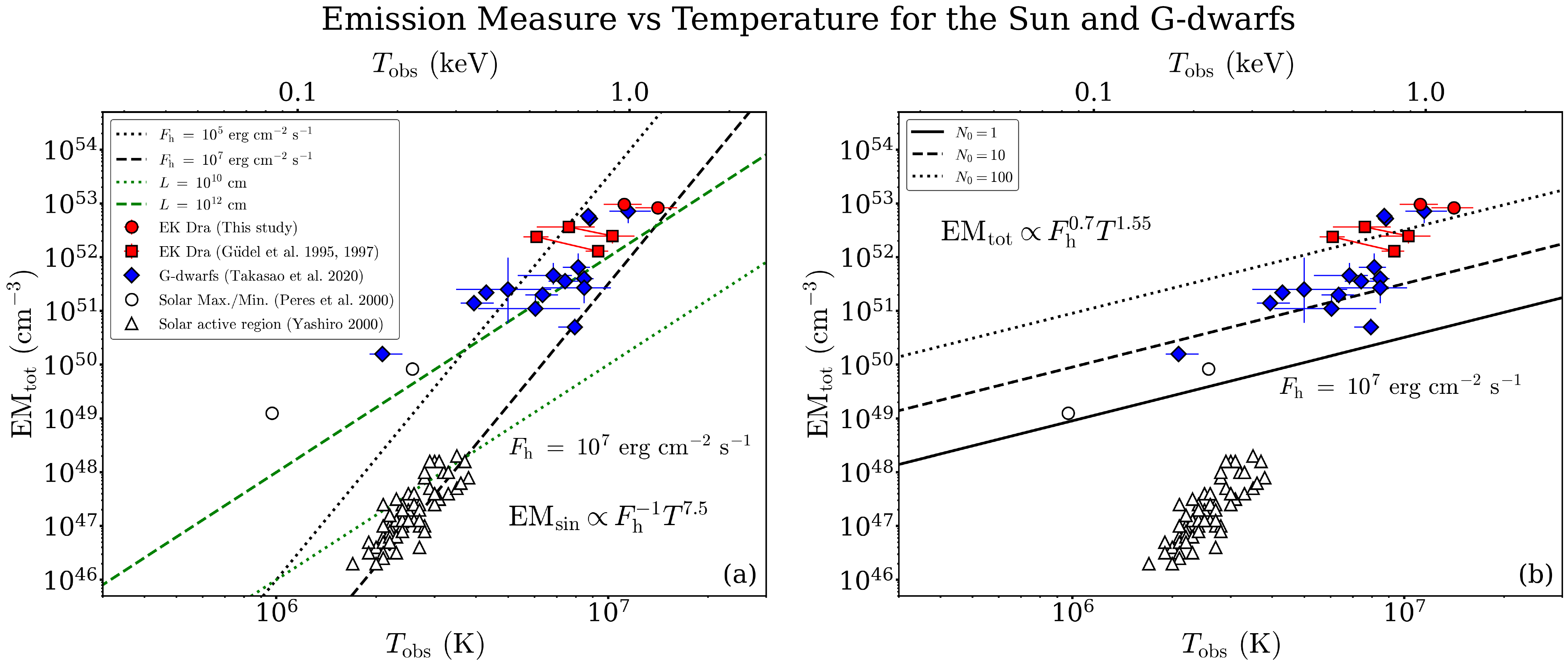}
\caption{The emission measure (EM) versus temperature for the Sun and various G-dwarfs. (Red) EK Dra in this study \citepalias[for details,][]{Namekata23} and the previous studies \citep{ Guedel95, Guedel97}. 
The EM and temperature are adopted with weighted and higher temperatures for the previous studies of EK Dra (left to right).
(Blue) G-dwarfs \citep{Takasao20} including $\pi^1$ Ursae Majoris, $\beta$ Hydri in \cite{Guedel97}, and $\iota$ Horologii in \cite{Sanz-Forcada19}. (Black) The maximum and minimum of the solar activity with multiple active regions \citep{Peres00} in \cite{Takasao20}, and single active regions \citep{Yashiro00} in \cite{Yashiro01} and \cite{Shibata02}.
(a) The typical EM from the heating flux or loop length is calculated with Equation \ref{eq:em_sin} (black and green). (b) The total EM is represented by ${\rm EM}_{\rm tot} \propto F_{\rm h}^{0.7} T^{1.55} $ under $\alpha=1.85$, and the typical EM is calculated for different number coefficients $N_0$ with the values of $f=0.1$, $\bar{B}=100 \ \text{G}$, and $F_{\rm h} = 10^7 \ \text{erg} \ \text{cm}^{-2} \ \text{s}^{-1}$ in Equation \ref{eq:em_tot} (black).
}\label{fig:EM_T}
\end{figure*}

In Figure \ref{fig:EM_T}, we compare the emission measure (EM) and temperature of EK Dra at the time of the pre-flare of E2 and E3 \citepalias[Table 5 in][]{Namekata23} with previous studies of EK Dra \citep{Guedel95,Guedel97}, various G-dwarfs, and the maximum and minimum of the solar activity \citep[][]{Peres00} in \cite{Takasao20}, and single active regions on the Sun \citep{Yashiro00} in \cite{Yashiro01} using the formula between the EM and temperature based on models of a single active region and multiple active regions (for details, Appendix \ref{sec:t_em}). 
As discussed in \cite{Takasao20}, the EM of the solar maximum/minimum and G-dwarfs cannot be reasonably explained by the model of a single active region \citep{Shibata02} in the left panel. The spatial scales for the EM are required to be close to $L = 10^{12}$ cm, which is larger than the solar radius ($R_{\rm Sun} = 6.96 \times 10^{10}$ cm). 
The EM should be explained by the model of multiple active regions under the different number coefficients in the right panel.
EK Dra is also located above the loop length longer than $L > 10^{12}$ cm if we assume the EM from the model of a single active region, but this value is fifteen times larger than the stellar radius ($R_{\rm star} = 0.94R_{\rm Sun} = 6.54 \times 10^{10}$ cm). This assumption is unlikely to explain the observed EM.
Thus, the EM of EK Dra should be explained by multiple active regions, as proposed for various G-dwarfs in \cite{Takasao20}. This can be consistent with the above suggestion on the existence of multiple spots from the multiwavelength variability.

\subsection{Correspondence between starspots and superflares}\label{sec:timing}
We compare the Light-curve and Doppler maps with superflares with and without prominence eruptions (E1, E2, and E3) at the times of 2680.0329, 2685.9980, and 2687.0119 (BJD - 2457000), respectively \citepalias{Namekata23}. This is because large superflares could originate from large spots and strong magnetic fields on the stellar surface \citepalias{Namekata24}.
In Table \ref{tb:prom}, we deduce the posterior distributions of spot longitude, colongitude, and radius at the times of E1, E2, and E3, from the posterior distributions of the spot mapping for both models. As in \citetalias{Namekata24}, we label the first, second, and third spots in the Light-curve map, as spot A, B, and C, respectively.
We also visualize the Light-curve and Doppler maps at different rotational phases including those of E1, E2, and E3 (Figure \ref{fig:compare_doppler}).

 As in \citetalias{Namekata24}, we discuss the Light-curve maps based on the three-spot model for the E1 in this study. Figure \ref{fig:S3_spotmap} shows that there are all three spots on mid-latitudes (25, 25, and 48 deg) at the time of E1 (Table \ref{tb:para}), and the spot B on the mid-latitude (25 deg) is suggested to be near the stellar limb (58 deg from the center) in the three-spot model (Table \ref{tb:prom}). The Doppler map and radial magnetic field suggest large dark regions and the PIL around the spot at the time of E1. Thus, \citetalias{Namekata24} suggests that the large prominence eruption E1 could originate from the spot B. This is supported by the two-spot model such that Figure \ref{fig:S2_spotmap} shows that the two spots are on the mid-latitude (51 deg) and equator (2 deg) at the time of E1, and the spot B is suggested to be on the stellar limb (83 deg from the center).

For the E2, it is suggested to be the spot A respectively on mid-latitudes (51 and 25 deg) in the stellar disk (37 and 29 deg from the center) for both models. These spot latitudes are different probably due to the degeneracy in the spot mapping, and almost near the dark region in the Doppler map and the PIL in the radial magnetic field within the spatial resolution of the ZDI ($\simeq$ 25 deg; see also Appendix B in \citetalias{Namekata24}). Thus, the prominence eruption E2 may be caused around the spot A for each of the model.
For the E3, it is suggested to be the spot B and C respectively for the two- and three-spot models on the latitude (2 and 48 deg) near the stellar limb (-49 and -59 deg from the center). These spots are also near the dark region in the Doppler map and on the radial field strength of $\sim$ 0 G. Thus, the superflare E3 may be caused by these spots for each of the model. 
To confirm the relation between starspots and superflares further, we intend to perform the DI and monitoring observation of superflares simultaneously in future works (Section \ref{sec:conclusion}).

\section{Conclusion and Future prospects} \label{sec:conclusion}
In \citetalias{Namekata23}, we reported on the discovery of large prominence eruptions associated with two of three superflares on a young solar-type star EK Dra through the multiwavelength observations in the optical, H$\alpha$ spectrum, and X-ray. The prominence eruptions were observed as H$\alpha$ emission, and the larger event was additionally accompanied by possible X-ray dimming interpreted as an evidence for a stellar CME.
In \citetalias{Namekata24}, we investigated the dynamics of the large prominence eruption using a data-driven modeling by the semi-analytical free-fall model and hydrodynamic simulations. We further demonstrated the magnetic field is likely to be the origin of the prominence eruption through the ZDI concurrently with the superflares, and performed starspot mapping from the TESS light curve based on this study. These results suggest that the prominence eruption was originated around spots on the PIL at mid-latitudes near the stellar limb.

In this study (Paper I\hspace{-.1em}I\hspace{-.1em}I), we perform starspot mapping for the TESS light curve of EK Dra and compare the result with the magnetic fields from the ZDI (Section \ref{sec:di}). We further explore multiwavelength variability ascribed to spots and active regions (Section \ref{sec:multiwave}), and investigate the relation between the spot distribution and superflares (Section \ref{sec:timing}). The result is summarized as follows:

\begin{itemize}
    \item[(i)] The spots are deduced to be on the equator to mid-latitude from the TESS light curve for two- and three-spot models. The spot locations (Light-curve map) are almost consistent with that in the intensity map obtained from the ZDI (Doppler map), except for a polar spot. This is because the polar spot gives a small contribution to the modulation of the light curve, and we assume only two- and three-spot models to reproduce the overall light curve, while a polar spot can be reconstructed from the ZDI.
    
    \item[(ii)] The H$\alpha$ light curve exhibits clear periodicity with respect to the TESS light curve as well as the Sun whereas the X-ray does not show such clear periodicity probably because of multiple spots on the surface and the spatial structure of the stellar corona. This can be supported by the relation between the emission measure and temperature from the X-ray spectrum.
    \item[(iii)] Either spot is near the stellar limb for large prominence eruption (E1) and superflare with an excess H$\alpha$ emission (E3), and on the stellar disk for the prominence eruption (E2), from both of the two- and three-spot models.
\end{itemize}

This study facilitates exploring starspots and magnetic fields of solar-type stars for the choromosphere and corona through multiwavelength observations in addition to the connection with plasma eruptions associated with superflares in \citetalias{Namekata23} and \citetalias{Namekata24}. 
EK Dra was observed in TESS Sector 75, 76, and 77 (January 30 to April 23 in 2024) through the TESS second extended mission (Cycle 6). Bright solar-type stars including EK Dra are to be observed in two-color photometry by the PLAnetary Transits and Oscillation of stars \citep[PLATO;][]{PLATO,Rauer25}. Solar-type stars are planned to be observed in X-ray and ultraviolet (XUV) by the Life-environmentology, Astronomy, and PlanetarY Ultraviolet Telescope Assembly \citep[LAPYUTA;][]{LAPYUTA} and infrared with by the Japan Astrometry Satellite Mission for INfrared Exploration \citep[JASMINE;][]{JASMINE}.

Based on this study, we intend to develop mapping the stellar surface with starspots and stellar faculae from multicolor light curves, taking into account the the intensity dependence of the faculae on the viewing angle \citep[e.g.,][]{Norris23}.
We can obtain more robust information of the surface map from multicolor photometry \citep[][]{Rosich20,Mori24} including spot temperature \citep[e.g.,][]{Waalkes23,Biagini24,Mori25} and because the contribution from the spot intensity and limb-darkening law to the light curve is different in multicolor photometry. Joint mapping from photometry and high-resolution spectroscopy (Doppler Imaging) also enables us to obtain more robust brightness map \citep[][]{Waite11,Finociety21,Finociety23,Lee25}. 
The time-dependent (Zeeman) Doppler Imaging technique is also required for the joint mapping of the stellar surface and magnetic fields \citep[e.g.,][]{Finociety22}. Robust mapping of the stellar surface also plays one of important roles in quantifying the effects of starspots and stellar faculae \citep{Norris23,Smitha25} on exoplanetary atmospheric characterizations \citep[][]{Rackham19,Rackham23,Thompson24, Berardo24, Rackham24} by the James Webb space telescope \citep[JWST;][]{jwst} and Ariel \citep{ariel}.

It is also important to investigate the relation between spots and activity indicators such as Balmer and Calcium lines through high-resolution spectroscopy \citep[e.g.,][]{Notsu13b, Nogami14, Notsu15}. 
In particular, we intend to utilize the High Dispersion Echelle Spectrograph \citep[HIDES;][]{HIDES} on the Okayama 1.88m telescope and the Gunma Astronomical Observatory Echelle Spectrograph for Radial Velocimetry \citep[GAOES-RV;][]{GAOES} with the MId Dispersion Spectrograph for Stellar Activity Research (MIDSSAR) on the 3.8m Seimei telescope \citep{Kurita20}.
We can obtain the H$\alpha$ spectrum and conduct the Doppler Imaging simultaneously through the spectroscopic monitoring observation of flare stars with the GAOES-RV and MIDSSAR to explore the dynamic descriptions of superflares from starspots to plasma eruptions associated with superflares.

\begin{acknowledgements}
We thank the anonymous referee for careful and constructive feedbacks which significantly improved both the content and clarity of this manuscript.
We appreciate Seiji Yashiro for providing data of the emission measure and temperature for solar active regions.
This study is based on publicly available data obtained by the TESS mission through the MAST data archive at the Space Telescope Science Institute (STScI). Funding for the
TESS mission is provided by NASA Science Mission Directorate.
The B-band photometric data were obtained by the 11.5cm telescope owned by H.M..
The optical spectroscopic data were
obtained through the program number 22A-N-CN06 (PI: K.N.) with the 3.8m Seimei telescope at Okayama Observatory, Kyoto University.
The X-ray data were obtained through the program IDs 5202490102 to 5202490113 (PI: V.S.A.) with the NICER on the ISS.
The spectropolarimetic data for the ZDI were obtained with the Neo-NARVAL on the TBL at the observatoire du Pic du Midi (BCool members: P.P., A.A.V., S.V.J., S.M., J.M., and C.N.).
Numerical computations were carried out on Yukawa-21 at the Yukawa Institute for Theoretical Physics, Kyoto University.
This work is supported by JSPS KAKENHI Grant Numbers JP20K04032, JP20H05643(H.M.), JP21H01131(H.M., D.N., and K.S.), JP21J00316(K.N.), JP22K03685(H.M.), JP23K17694(K.S.), JP24H00248(K.I., K.N., H.M., and D.N.), JP24K00680(K.N., H.M., D.N., and K.S.), JP24K00685(H.M.), JP24K17082(K.I.), and	JP25K01041(K.N. and H.M.). V.S.A. was supported by the GSFC Sellers Exoplanet Environments Collaboration (SEEC), which is funded by the NASA Planetary Science Division’s Internal Scientist Funding Model (ISFM), NASA's Astrophysics Theory Program grant \#80NSSC24K0776NNH21ZDA001N-XRP, F.3 Exoplanets Research Program grants, NICER Cycle 2 project funds and NICER DDT program.
A.A.V. acknowledges funding from the European Research Council (ERC) under the European Union's Horizon 2020 research and innovation programme (grant agreement No 817540, ASTROFLOW) and from the Dutch Research Council (NWO), with project number VI.C.232.041 of the Talent Programme Vici.
Y.N. was supported from the NASA ADAP award program Number 80NSSC21K0632,
NASA TESS Cycle 6 Program 80NSSC24K0493, and NASA NICER Cycle 6 Program 80NSSC24K1194.
S.V.J. acknowledges the support of the DFG priority program SPP 1992 ``Exploring the Diversity of Extrasolar Planets" (JE 701/5-1).
\end{acknowledgements}

\facilities{TESS, Seimei (KOOLS-IFU), NICER, TBL (Neo-NARVAL)}

\software{\texttt{eleanor} \citep{Feinstein19}; 
\texttt{stella} \citep{Feinstein20};
\texttt{tinygp} \citep{tinygp};
\texttt{JAX} \citep{jax};
\texttt{NumPyro} \citep{numpyro};
}

\appendix

\begin{deluxetable*}{lcc}[tbhp!]
\tablecaption{EK Dra (Sector 48, 49, and 50)
\label{tb:para_all}}
\tabletypesize{\scriptsize}
\tablehead{
\colhead{Deduced Parameters} &   \colhead{Three-spot Model} & \colhead{Prior Distribution\tablenotemark{{\rm a}}}  
}
\startdata
(Stellar parameters) && \\
1.  Equatorial period $P_{{\scriptsize \textrm{eq}}}$ (day)  &$ 2.5906 ^{+ 0.0001}_{- 0.0001 }$&${\cal N}_{\log} (2.51,0.08^2)$\tablenotemark{{\rm b}}\\
2.   Degree of differential rotation $\kappa$  & $ 0.0698 ^{+ 0.0001 }_{- 0.0001 }$&${\cal N} (0.108,0.104^2)$\tablenotemark{{\rm b}}\\
(Spot parameters) &  &    \\
(First spot) &  &    \\
3.     Latitude $\Phi_1$  (deg)&$ 23.02 ^{+ 0.02}_{- 0.04 }$&{${\cal U} (\Phi_2,\Phi_3)$\tablenotemark{{\rm c}}}\\
4.     Initial longitude $\Lambda_1$ (deg)  &$ 82.46 ^{+ 0.01}_{- 0.02}$ &${\cal U}  (-180.00,180.00)$\\
5.     Reference time $t_{1}$ (day)  &$ 2.227 ^{+ 0.018 }_{- 0.019 }$&${\cal U} (-57.341,26.229)$\\
6.     Maximum radius $\alpha_{{ \textrm{max,1}}}$ (deg)&$ 7.100 ^{+ 0.001 }_{- 0.001 }$ &${\cal U}(0.01,20.00)$\\
7.     Emergence duration ${\cal I}_1$ (day) &$ 999.996^{+ 0.004}_{- 0.154}$
&${\cal U}_{\log} (0.000,1000.000)$\\
8.     Decay duration ${\cal E}_1$ (day)&$ 999.999 ^{+0.001}_{- 0.149}$
 &${\cal U}_{\log} (0.000,1000.000)$\\
 (Second spot)&  &     \\
9.     Latitude $\Phi_2$  (deg) &$ -25.85 ^{+ 0.03 }_{- 0.02 }$ &{${\cal U} (-90.00,\Phi_1)$\tablenotemark{{\rm c}}}\\
10.     Initial longitude $\Lambda_2$  (deg)& $ -141.96 ^{+ 0.01 }_{- 0.02 }$&${\cal U}  (-180.00,180.00)$\\
11.     Reference time $t_{2}$ (day) &$ -57.285 ^{+ 0.067 }_{- 0.001 }$&${\cal U} (-57.341,26.229)$ \\
12.     Maximum radius $\alpha_{{\textrm{max,2}}}$  (deg)&$ 10.512 ^{+ 0.006 }_{- 0.006 }$&${\cal U} (0.01,20.00)$\\
13.     Emergence duration ${\cal I}_2$ (day) &$ 121.945 ^{+ 16.248 }_{- 27.887 }$ &${\cal U}_{\log} (0.000,1000.000)$\\
14.     Decay duration ${\cal E}_2$ (day) &$ 829.153^{+ 4.669 }_{- 2.301 }$ &${\cal U}_{\log} (0.000,1000.000)$\\
(Third spot) &    & \\
15.     Latitude $\Phi_3$  (deg)  & $ 71.40 ^{+ 0.06 }_{- 0.02 }$
 &${\cal U} (\Phi_2,90.00)$\tablenotemark{{\rm c}}\\
16.     Initial longitude $\Lambda_3$ (deg)  & $ -107.78 ^{+ 0.01}_{- 0.02}$
&${\cal U}  (-180.00,180.00)$\\
17.     Reference time $t_{3}$ (day)  &$ 22.259 ^{+ 0.005 }_{- 0.006 }$
 &${\cal U} (-57.341,26.229)$ \\
18.     Maximum radius $\alpha_{{ \textrm{max,3}}}$ (deg) &$ 10.514 ^{+ 0.013 }_{- 0.006 }$&${\cal U}(0.01,20.00)$\\
19.      Emergence duration ${\cal I}_3$ (day) &$ 320.476 ^{+ 0.610 }_{- 0.350 }$ &${\cal U}_{\log} (0.00,1000.000)$\\
20.     Decay duration ${\cal E}_3$ (day) &$ 19.594 ^{+ 0.060 }_{- 0.048 }$ &${\cal U}_{\log} (0.00,1000.000)$\\
\hline
(Derived values) & &   \\
Equatorial rotational velocity $\Omega_{\rm eq}$ (rad day$^{-1}$) &$ 2.4254 ^{+ 0.0001 }_{- 0.0001 }$ & \\
Rotational shear $\Delta \Omega$ (rad day$^{-1}$)  &$ 0.1692 ^{+ 0.0002 }_{- 0.0001 }$ & \\ \hline
 & $1.0140^{+0.0001}_{-0.0001}$ & \\
Unspotted level\tablenotemark{{\rm d}} & $1.0160^{+0.0001}_{-0.0001}$ & \\
& $1.0174^{+0.0001}_{-0.0001}$ & \\ \hline
Reduced chi-square & $3793.941^{+0.007}_{ -0.010}$ &  \\ 
Logarithm of Model evidence $\log{\cal Z}$ & -4145577.491 & \\ \hline
\enddata
\tablenotetext{{\rm a}}{
${\cal U_{\text{log}}}(a,b)= 1/(\theta \log(b/a))$, ${\cal U}(a,b)=1/(b-a)$, ${\cal N}(\mu,\sigma^2)$, and ${\cal N}_{\rm log}(\mu,\sigma^2) = (1/\theta) \times {\cal TN}(\log \mu,(\sigma/\mu)^2)$  represent the bounded log uniform distribution (Jeffreys prior distribution) and bounded uniform distribution defined in $a\leq \theta \leq b$, and normal distribution and log-normal distribution with the mean $\mu$ and standard division $\sigma$ defined in $0 \leq \theta $, respectively.
}
\tablenotetext{{\rm b}}{The parameters in the spotted model correspond to $P_{\rm eq} = 2 \pi/ \Omega_{\rm eq}$ and $\kappa = \Delta \Omega/  \Omega_{\rm eq}$ \citep[Equation 5 in][]{Ikuta23}. We set normal prior distributions for $P_{\rm eq}$ and $\kappa$ with an error propagation in Table \ref{tb:stellar}.
}
\tablenotetext{{\rm c}}{
We discern each spot by its latitude $\Phi_k$ as in \cite{Ikuta23}. In the case of the two-spot model, we set $\Phi_3 = 90.0$ (the upper limit of the latitude).
}
\tablenotetext{{\rm d}}{
The unspotted levels take different values respectively for Sector 48, 49, and 50, because we adopt temporally variable spot radius and each of the time-averaged projected area during a sector is different.
}
\end{deluxetable*}

\begin{deluxetable*}{lccc}[tbhp!]
\tablecaption{Spot parameters from Sector 48, 49, and 50, at the time of Event 1, 2, and 3 \label{tb:prom_all}}
\tabletypesize{\scriptsize}
\tablehead{
\colhead{Parameters} &  \colhead{Event 1}  &  \colhead{Event 2}  & \colhead{Event 3}
}
\startdata
(Spot A)& &  &  \\
Longitude   (deg)$^{\rm a}$&$ -48.78 ^{+ 0.02 }_{- 0.02}$   & $ 51.30 ^{+ 0.02 }_{- 0.03 }$  & $ -169.31 ^{+ 0.02 }_{- 0.02}$\\
Colongitude (deg)$^{\rm b}$&$ -43.85 ^{+ 0.02 }_{- 0.03}$ &  $ 45.99 ^{+ 0.02 }_{- 0.02}$ &$ -163.80^{+ 0.03 }_{- 0.05}$   \\ 
Radius  (deg) &$ 7.011 ^{+ 0.001 }_{- 0.001 }$ &$ 6.969 ^{+ 0.001 }_{- 0.001 }$  &$ 6.962 ^{+ 0.001}_{- 0.001 }$ \\
Radius   ($10^{10}$ cm) & $ 0.79 ^{+ 0.07 }_{- 0.05 }$  &$ 0.78^{+ 0.07}_{- 0.05 }$  & $ 0.80 ^{+ 0.05 }_{- 0.07 }$\\
(Spot B)& &  &  \\
Longitude (deg)$^{\rm a}$&$ 81.48 ^{+ 0.02 }_{- 0.02}$&$ 179.41 ^{+ 0.02}_{- 0.02}$ & $ -41.56 ^{+ 0.02 }_{- 0.02}$\\ 
Colongitude (deg)$^{\rm b}$&$ 96.57^{+ 0.02}_{- 0.03}$ &$ 179.47^{+ 0.02 }_{- 0.02 }$ & $ -58.55 ^{+ 0.02 }_{- 0.02}$ \\ 
Radius  (deg) &$ 9.601 ^{+ 0.005}_{- 0.006 }$  &$ 9.525 ^{+ 0.005 }_{- 0.005 }$  &$ 9.513 ^{+ 0.004}_{- 0.006 }$ \\
Radius  ($10^{10}$ cm) &$ 1.08^{+ 0.10 }_{- 0.07 }$ &$ 1.09 ^{+ 0.08 }_{- 0.08 }$ &$ 1.09 ^{+ 0.08}_{- 0.08 }$\\
(Spot C)& &  &  \\
Longitude  (deg)$^{\rm a}$ &$ 14.27^{+ 0.02 }_{- 0.02 }$&$ 71.21^{+ 0.02}_{- 0.02}$&$ -156.73 ^{+ 0.02 }_{- 0.02}$\\
Colongitude (deg)$^{\rm b}$&$ 6.04 ^{+ 0.01 }_{- 0.01}$ &$ 28.21 ^{+ 0.03}_{- 0.07 }$ &  $ -29.79 ^{+ 0.18 }_{- 0.06}$ \\ 
Radius  (deg) &$ 10.268 ^{+ 0.013 }_{- 0.006 }$&$ 10.464 ^{+ 0.013 }_{- 0.006 }$  & $ 10.497 ^{+ 0.013 }_{- 0.006 }$\\
Radius ($10^{10}$ cm) &$ 1.17 ^{+ 0.09 }_{- 0.09 }$&$ 1.19^{+ 0.10 }_{- 0.08 }$ & $ 1.20 ^{+ 0.09 }_{- 0.09 }$\\
\enddata
\tablenotetext{{\rm a}}{
The longitude for each of the spot latitude $\Phi_k$ at the time $t$ is calculated by $\Lambda_k + 2 \pi t/P(\Phi_k)$ (deg), where $P(\Phi_k)$ is given by  Equation \ref{eq:rot}. The radius at the time $t$ is calculated by Equation \ref{eq:spotrad} and transformed by multiplying the stellar radius in Table \ref{tb:stellar}.}
\tablenotetext{{\rm b}}{The tangent of the colongitude is given by $\sin \Lambda_t \cos \Phi/(\cos i \sin \Phi + \cos \Phi \sin i \cos \Lambda_t)$, where $\Lambda_t = \Lambda + 2 \pi t/P(\Phi)$ at the time of $t$.}
\end{deluxetable*}

\begin{figure*}[tbhp!]
\plotone{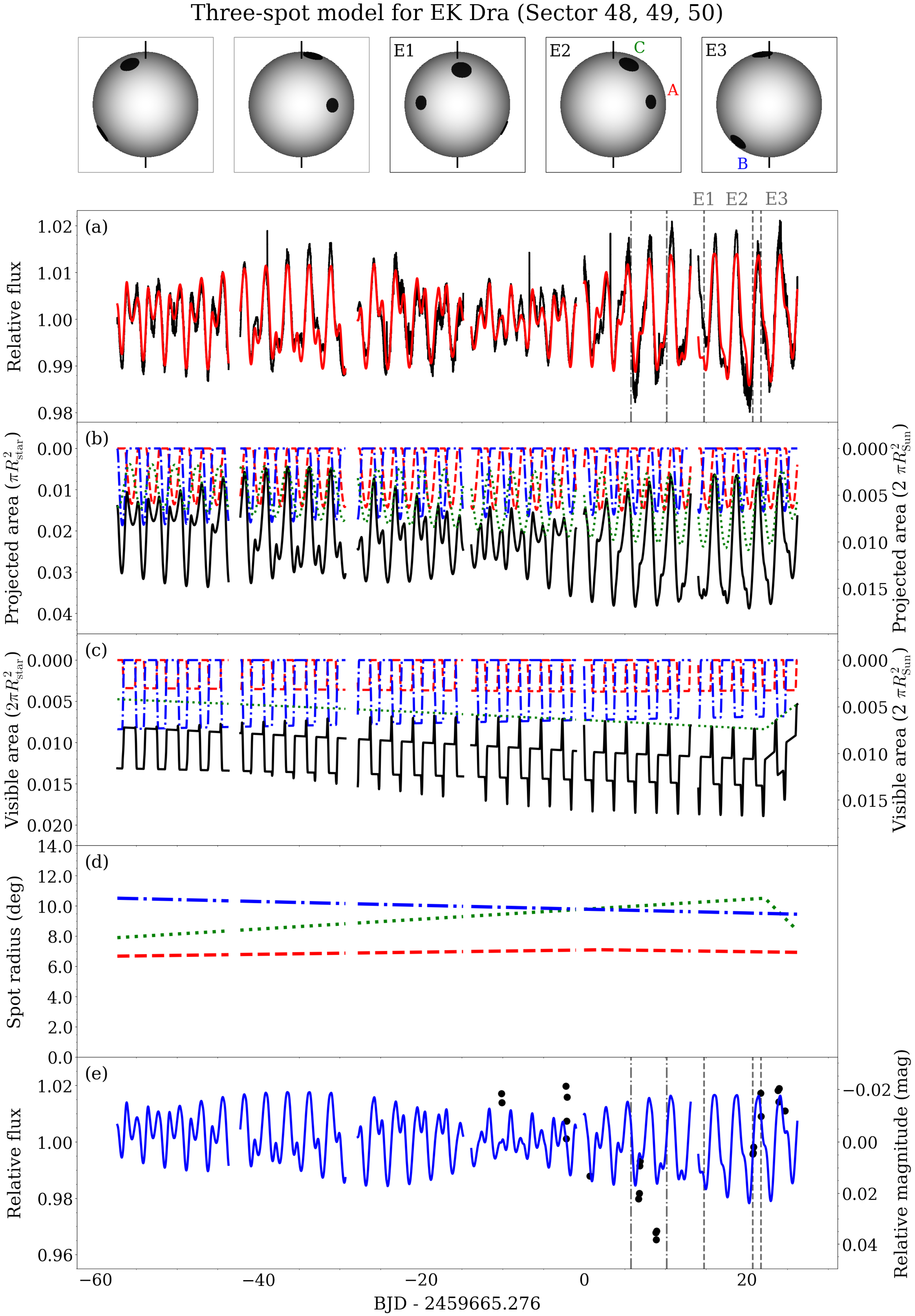} 
\caption{Same as Figure \ref{fig:S3_spotmap}, but for the TESS light curve in the Sector 48, 49, and 50.
} \label{fig:spotmap_all}
\end{figure*}

\section{TESS data in consecutive three sectors} \label{sec:appendix}
EK Dra was also observed in the Sectors 48 and 49 (January 28 to March 26 in 2022), while we use the TESS data only in the Sector 50 (Section \ref{sec:tess}). In Appendix \ref{sec:appendix}, we use the TESS data in these consecutive three Sectors. We retrieve the data in Sectors 48 and 49 in the same manner for the Sector 50. We extract flare candidates from the retrieved PCA-flux by comparing detected ones in the PDC-SAP (Pre-search Data Conditioning Simple Aperture Photometry) flux using \texttt{stella} \citep{Feinstein20}. 
We perform starspot mapping only with the three-spot model and reproduce the B-band light curve (Section \ref{sec:spotmap} and \ref{sec:bband}). We note that we adopted the two-spot model, but the two-spot model cannot reproduce the overall of the TESS light curve.

Table \ref{tb:para_all} and \ref{tb:prom_all} exhibit the deduced parameters, and Figure \ref{fig:spotmap_all} exhibits the visualized surface, the TESS and reproduced light curves, temporal variation of the spot area, and reproduced B-band light curve.
The derived parameters are different especially for the spot latitude and radius because we performed starspot mapping with only three spots, and the result of the starspot mapping depends on the assumed model \citep{Basri20,Luger21a}.
The TESS light curve can be well reproduced even with the three-spot model. 
In this case, always visible spot is deduced to be on the latitude of 71 deg (spot C), which may be consistent with the polar spot from the ZDI.

\section{Coronal properties} \label{sec:t_em}
We investigate the stellar coronal properties using the relation between the emission measure (EM) and temperature \citep{Shibata02,Takasao20,Notsu25} in Section \ref{sec:multiwave}. In Appendix \ref{sec:t_em}, we describe the formula between the EM and temperature for active regions \citep[for details,][]{Takasao20}.

The EM from a single active region is formulated by
\begin{align} \label{eq:em_sin}
\text{EM}_{\rm sin} &\simeq 10^{44} (\frac{f}{0.1}) (\frac{T}{10^6 \ \text{K}})^{15/2}  (\frac{F_{\rm h}}{10^7 \ \text{erg} \ \text{cm}^{-2} \ \text{s}^{-1}})^{-1} \ {\rm cm}^{-3} \notag \\
&\simeq 10^{46} (\frac{f}{0.1})  (\frac{T}{10^6 \ \text{K}})^{4} 
 (\frac{L}{10^{10} \ {\rm cm}}) \ {\rm cm}^{-3},
\end{align}
where $f=0.1$ is the filling factor of the coronal active region, $T$ is the temperature of the active region, $F_{\rm h}$ is the heating flux, and $L$ is the typical length of the active region, respectively.
The total EM with multiple spots is also derived based on the size distribution of sunspots:
\begin{align} \label{eq:em_tot}
{\rm EM}_{\rm tot} &\simeq \frac{14 N_0 \gamma_0}{29-14 \alpha} 10^{60-16\alpha} (\frac{f}{0.1}) \notag \\ &\times (\frac{F_{\rm h}}{10^7 \ \text{erg} \ \text{cm}^{-2} \ \text{s}^{-1}})^{-3+2\alpha}(\frac{T_{\rm max}}{10^6 \ \text{K}})^{(29-14\alpha)/2} \ {\rm cm}^{-3},
\end{align}
where $N_0$ is the number coefficient relative to the size distribution of sunspots, $\gamma_0=1.86 \times 10^{19} \bar{B}^{-\alpha+1}$ is the coefficient dependent on the strength of the magnetic field $\bar{B}$, $\alpha=1.85$ is the power-law index of the size distribution, and $T_{\rm max}$ is the maximum temperature of active regions, respectively. We set $\bar{B}=100 \ \text{G}$ as the typical strength of the magnetic field.
The observed temperature can be also represented by $T_{\rm obs} = (29-14\alpha)/(31-14\alpha) T_{\rm max} \simeq 0.61 T_{\rm max}$.

Figure \ref{fig:EM_T} shows the emission measure (EM) versus temperature for the Sun and various G-dwarfs: EK Dra in this study and previous studies \citep{Guedel95, Guedel97}, G-dwarfs \citep{Takasao20} including $\pi^1$ Ursae Majoris, $\beta$ Hydri in \cite{Guedel97}, and $\iota$ Horologii in \cite{Sanz-Forcada19}, the maximum and minimum of the solar activity with multiple active regions \citep{Peres00} in \cite{Takasao20}, and single active regions on the Sun \citep{Yashiro00} in \cite{Yashiro01} and \cite{Shibata02}.
In this study, because the X-ray spectra of EK Dra is fitted with four components in \citetalias{Namekata23}, we calculate the observed temperature weighted with the EM and take the summation of the EM.
For EK Dra in \cite{ Guedel95, Guedel97}, we show both values for weighted and higher temperatures. The EM and temperature of EK Dra in the previous studies are estimated to be lower than those in this study because of lower range of the X-ray wavelength by the ROSAT (0.1 to 2.4 keV) than that of the NICER (0.5 to 3 keV) (Section \ref{sec:nicer}).
For the Sun and other G-dwarfs in \cite{Takasao20}, we adopt the component with higher temperature as the observed EM and temperature if the X-ray spectra is fitted with two components. 
The solar maximum and minimum \citep{Peres00} have larger EM than single active regions \citep{Yashiro00}. This is because the EM of the solar maximum and minimum is summed up with multiple active regions and observed especially in the energy range sensitive to the typical coronal temperature (0.1 to 0.3 keV) by the ROSAT/PSPC (0.1 to 3 keV) while the single active regions are observed by the Yohkoh/SXT (AlMg filter: 0.3 to 5 keV).

\newpage

\bibliography{ekdra}{}


\bibliographystyle{aasjournal}

\end{document}